\documentclass[aps,superscriptaddress,preprint,nofootinbib,eqsecnum]
{revtex4}
\usepackage{amsmath,amssymb,bm}
\usepackage{graphicx}

\def\be{\begin{equation}}
\def\ee{\end{equation}}

\def\Vol{\mbox{Vol}}

\def\tr{\mbox{tr}\,}

\def\k{{\bm k}}

\def\CA{{\cal A}}
\def\CB{{\cal B}}
\def\qn{q}
\def\wn{w}
\def\eps{\varepsilon}
\newcommand\gfourD{g_{\mathrm{4D}}}
\renewcommand\d{\partial}
\renewcommand\Im{\mathop{\mathrm{Im}}}

\begin{document}
  \vspace*{-1.5cm}
\begin{flushright}
  {hep-th/0701036}\\
  {NSF-KITP-06-129}\\
  {INT-PUB 07-01}
  \end{flushright}
  \vspace*{1cm}

\title{Quantum critical transport, duality, and M-theory}

\author{Christopher P. Herzog}
\affiliation{Department of Physics, University of Washington,
Seattle, WA 98195-1560, USA}

\author{Pavel Kovtun}\affiliation{Kavli Institute for Theoretical
Physics,
University of California, Santa Barbara, CA 93106-4030, USA}

\author{Subir Sachdev}
\affiliation{Department of Physics, Harvard University, Cambridge MA
02138, USA}

\author{Dam Thanh Son}
\affiliation{Institute for Nuclear Theory, University of Washington,
              Seattle, WA 98195-1550, USA}

\begin{abstract}
\noindent We consider charge transport properties of 2+1
dimensional conformal field theories at non-zero
temperature. For theories with only Abelian U(1) charges, we
describe the action of particle-vortex duality on the
hydrodynamic-to-collisionless crossover function:
this leads to powerful functional constraints for self-dual theories.
  For
$\mathcal{N}{=}8$ supersymmetric, SU($N$) Yang-Mills theory at the
conformal fixed point, exact hydrodynamic-to-collisionless crossover
functions of the SO(8) R-currents can be obtained in the large $N$
limit by applying the AdS/CFT correspondence to M-theory. In the
gravity theory,
fluctuating currents are mapped to fluctuating gauge fields in the
background of a black hole in 3+1 dimensional anti-de Sitter space.
The electromagnetic self-duality of the 3+1 dimensional theory
implies that the correlators of the R-currents obey a functional
constraint similar to that found from particle-vortex duality in 2+1
dimensional Abelian theories. Thus the 2+1 dimensional,
superconformal Yang Mills
theory obeys a ``holographic self duality'' in the large $N$ limit,
and perhaps more generally.
\end{abstract}

\date{March 2007}

\maketitle

\section{Introduction}
\label{sec:intro}

The quantum phase transitions of two (spatial) dimensional systems
have been the focus of much study in the condensed matter community.
Prominent examples include the superfluid-insulator transition in
thin films \cite{gruner,kapi,shahar}, the transitions between
various quantum Hall states \cite{shahar2,engel}, and magnetic
ordering transitions of Mott insulators and superconductors which
have applications to the cuprate compounds
\cite{collin,dsz,younglee}. Of particular interest in this paper are
the transport properties of conserved quantities such as the
electrical charge or the total spin: these are characterized by a
(charge or spin) conductivity $\sigma$, which can in general be a
complicated function of frequency $\omega$, wavevector $k$,
temperature $T$, and various couplings characterizing the ground
state.

It is often the case that the quantum critical point is described by
a strongly interacting quantum field theory in 2+1 spacetime
dimensions $D$. Examples are ({\em i\/}) the superfluid-insulator
transition in the boson Hubbard model at integer filling
\cite{fwgf,bloch,spielman}, which is described by the $\varphi^4$
field theory with O(2) symmetry, and so is controlled by the
Wilson-Fisher fixed point in $D=2{+}1$; ({\em ii\/}) the spin-gap
paramagnet to N\'eel order transition of coupled spin
dimers/ladders/layers which is described by the O(3) $\varphi^4$
field theory \cite{wang,matsumoto}; and ({\em iii\/}) the
`deconfined' critical point of a $S=1/2$ antiferromagnet between a
N\'eel and a valence bond solid state \cite{senthil,anders}, which
is described by the ${\mathbb C}{\mathbb P}^1$ model with a
non-compact U(1) gauge field \cite{mv}. In all these cases the
critical point is described by a relativistic conformal field theory
(CFT). With an eye towards such experimentally motivated
applications, our purpose here is to explore the transport
properties of general interacting CFTs in $D=2{+}1$.

A crucial property of CFTs in $D=2{+}1$ (which actually applies more
generally to any critical theory in 2 spatial dimensions which obeys
hyperscaling) is that the conductivity is $1/\hbar$ times a
dimensionless number. For U(1) currents, there is also a prefactor of
$(e^{\ast})^2$ where $e^\ast$ is the unit of charge --- we will drop
this factor below. For non-Abelian Noether currents, the
normalization of charge is set by a conventional normalization of
the generators of the Lie algebra. We will be working with
relativistic theories, and therefore set $\hbar = k_B = c = 1$.

Initial discussions \cite{mpaf1,mpaf2,wenzee} of this dimensionless
conductivity at the quantum critical point were expressed in terms
of ground state correlations of the CFT. Let $J^a_{\mu}$ represent
the set of conserved currents of the theory; here $\mu =0,1,2$ is a
spacetime index, and $a$ labels the generators of the global
symmetry. In the CFT, $J^a_{\mu}(x)$ has dimension 2, and so current
conservation combined with Lorentz and scale invariance imply for
the Fourier transform of the retarded correlator
$C_{\mu\nu}^{ab}(x)$ 
at zero temperature%
\footnote{If needed, a ``diamagnetic'' or ``contact''
    term has been subtracted to ensure current conservation.
    In theories with Chern-Simons terms, an additional term
    proportional to $\epsilon_{\mu\nu\lambda} p_\lambda$ is permitted in
    Eq.~(\ref{j0}) and the $T>0$ generalization in Eq.~(\ref{j1}).
    See Appendix~\ref{app:cs}.
}:
\begin{equation}
C_{\mu\nu}^{ab}(p)\,\big|_{T{=}0} = \sqrt{p^2} \left( \eta_{\mu\nu}
- \frac{p_\mu p_\nu}{p^2} \right) K_{ab}\,, \label{j0}
\end{equation}
where $\eta_{\mu\nu}={\rm diag}(-1,1,1)$, $p_\mu=(-\omega,\k)$ is
spacetime momentum, and $p^2=\k^2-\omega^2$. We define $\sqrt{p^2}$
so that
it is analytic in the upper-half-plane of $\omega$ and  $\Im
\sqrt{p^2} \leq 0$ for $\omega>0$. The parameters $K_{ab}$
are a set of universal, momentum-independent dimensionless constants
characterizing the CFT, which are the analog of the central charge
of the Kac-Moody algebra of CFTs in $D=1{+}1$. Application of the
Kubo formula at $T{=}0$ shows that \cite{mpaf1,wenzee} the $K_{ab}$
are equal to the conductivities $\sigma_{ab} = K_{ab}$, thus setting
up the possibility of observing these in experiments.

It was also noted \cite{mpaf2,wenzee} that particle-vortex duality
\cite{peskin,dasgupta,mpaf3} of theories with Abelian symmetry
mapped the $T=0$ conductivities to their inverse (we review this
mapping in Section~\ref{sec:cp1}). In self-dual theories, this
imposes constraints on the values of the $K_{ab}$, possibly allowing
them to be determined exactly. However, the field theories
considered in these early works were not self-dual (see
Appendix~\ref{app:cs}). Duality, and possible self-duality, was also
considered in the context of theories containing Chern-Simons terms,
relevant to quantum Hall systems
\cite{leefisher,fradkin,pryadko,shimshoni,burgess,witten}. We comment
on these
works in Appendix~\ref{app:cs}, but the body of the paper considers
only theories without Chern-Simons terms. For our purposes, more
relevant is the self-dual field theory proposed recently by
Motrunich and Vishwanath \cite{mv}, and we discuss its charge
transport properties below.

It was subsequently pointed out \cite{damle,ssqhe,ssbook} that the
$K_{ab}$ are {\em not} the d.c.~conductivities observed at small but
non-zero temperature. The key point \cite{ssye,ssbook,hod} is that
at non-zero $T$, the time $1/T$ is a characteristic `collision' or
`decoherence' time of the excitations of the CFT. Consequently the
transport at $ \omega \ll T$ obeys `collision-dominated'
hydrodynamics, while that at $\omega \gg T$ involves `collisionless'
motion of excitations above the ground state. Therefore, the limits
$\omega \rightarrow 0$ and $T \rightarrow 0$ do not, in general,
commute, and must be taken with great care; the constants $K_{ab}$
above are computed in the limit $\omega /T \to\infty$, while the
d.c.~conductivities involve $\omega /T \to 0$.

This contrast between the collisionless and collision-dominated
behavior is most clearly displayed in the correlations of the
conserved densities. Taking the $tt$ component of Eq.~(\ref{j0}) we
obtain the response
\begin{equation}
C_{tt}^{ab}(\omega, k) = K_{ab} \frac{-k^2}{\sqrt{k^2 -
\omega^2}}~~~,~~~||\omega| - k| \gg T \,,\label{j0n}
\end{equation}
which characterizes the `collisionless' response of the CFT at
$T=0$. We have also noted above that we expect the same result to
apply at $T>0$ provided $\omega$ and $k = |\k|$ are large enough,
and away from the light cone. The $T>0$ correlations are the Fourier
transform of the retarded real time correlators. These are related
by analytic continuation to the Euclidean space correlations defined
at the Matsubara frequencies, which are integer multiples of $2 \pi
T$. The low frequency hydrodynamic regime $\omega \ll T$ is only
defined in real time (Minkowski space). In this regime, the
arguments of Ref.~\cite{damle} imply that the
`collision-dominated' response has the structure%
\begin{equation}
C_{tt}^{ab}(\omega, k) = \sum_{\lambda} \chi^\lambda_{ab}
\frac{-D_\lambda k^2}{-i \omega + D_\lambda k^2}~~~,~~~|\omega|, k
\ll T \,,\label{j0d}
\end{equation}
where $D_\lambda$ are the diffusion constants of a set of diffusive
eigenmodes labelled by $\lambda$, and $\chi^\lambda_{ab}$ are the
corresponding susceptibilities. Scaling arguments imply that
\cite{chubukov} $D_\lambda = \mathcal{D}_\lambda /T$ and
$\chi^\lambda_{ab} = \mathcal{C}^\lambda_{ab} T$, where the $
\mathcal{D}_\lambda, \mathcal{C}^\lambda_{ab}$ are a set of
universal numbers characterizing the hydrodynamic response of the
CFT. The d.c. conductivities can be obtained from the Kubo formula
by $\sigma_{ab} = \lim_{\omega \rightarrow 0} \lim_{k \rightarrow 0}
(i \omega/k^2) C_{tt}^{ab}$, where the order of limits is
significant. At any fixed $T>0$, the limits of small $k$ and
$\omega$ imply that this Kubo formula has to be applied to
Eq.~(\ref{j0d}), and leads to Einstein relations between the
$T$-independent universal conductivities and the diffusivities. The
distinct forms of Eqs.~(\ref{j0n}) and (\ref{j0d}) make it clear
that, in general, the universal d.c. conductivities bear no direct
relationship to the $K_{ab}$; the latter, as we will see below in
Eq.~(\ref{sigma0}), are related to the high frequency conductivity.

It is worth noting here in passing that the structure in
Eq.~(\ref{j0d}) does {\em not\/} apply to CFTs in $D=1{+}1$, where
a result analogous to
Eq.~(\ref{j0n}) holds also in the low frequency and low momentum
limit; see Appendix~\ref{app:cft2} for further discussion of this
important point.

Returning to consideration of all the components of the
$C_{\mu\nu}^{ab}$ in $D=2{+}1$, an alternative presentation of the
collisionless-to-hydrodynamic crossover is obtained by writing down
the generalization of Eq.~(\ref{j0}) to $T>0$. Current conservation
and spatial rotational invariance, without Lorentz invariance at
$T>0$, generalize Eq.~(\ref{j0}) to
\begin{equation}
   C_{\mu\nu}^{ab}(\omega , \k)
   = \sqrt{p^2} \Bigl( P^T_{\mu\nu}\, K^T_{ab} (\omega, k)
   + P^L_{\mu\nu}\, K^L_{ab} (\omega, k) \Bigr)
\label{j1}
\end{equation}
where $k = |\k|$, and $P^T_{\mu\nu}$ and $P^L_{\mu\nu}$ are
orthogonal projectors defined by
\begin{equation}
P^T_{00} = P^T_{0i} = P^T_{i0}=0~~,~~P^T_{ij} = \delta_{ij} -
\frac{k_i k_j}{k^2}~~,~~P^L_{\mu\nu} =
   \Big(\eta_{\mu\nu} - \frac{p_\mu p_\nu}{p^2}\Big) - P^T_{\mu\nu},
\end{equation}
with the indices $i,j$ running over the 2 spatial components. The
constants $K_{ab}$ have each been replaced by {\em two}
dimensionless, universal, temperature-dependent functions
$K_{ab}^{L,T}(\omega,k)$, characterizing the longitudinal and
transverse response. These functions are dimensionless, and
hence they can {\em only\/} depend upon the dimensionless ratios
$\omega/T$ and $k/T$, as is also the case for the conductivities.
Spatial rotational invariance, and the existence of finite
correlation length at $T>0$ which ensures analyticity at small $\k$,
imply that the longitudinal and transverse response are equal to
each other at $\k=0$, and, by the Kubo formula, are both equal to
the zero momentum, frequency dependent complex conductivity,
$\sigma_{ab} (\omega/T)$:
\begin{equation}
   \sigma_{ab}(\omega/T) = K_{ab}^L(\omega,0) = K^T_{ab}(\omega,0).
\label{sigmak}
\end{equation}
Also at $T=0$, these functions reduce to the constants in
Eq.~(\ref{j0}):
\begin{equation}
\sigma_{ab}(\infty)= K_{ab} = K_{ab}^L(\omega,k)\,\big|_{T{=}0} =
K_{ab}^T(\omega,k)\,\big|_{T{=}0}. \label{sigma0}
\end{equation}

The functions $K_{ab}^{L,T}(\omega,k)$ are clearly of great physical
interest, and it would be useful to compute them for a variety of
CFTs. A number of computations have appeared
\cite{damle,ssqhe,eric,sondhi1,sondhi2,sondhi3}, and show
interesting structure in the conductivity as a function of
$\omega/T$, encoding the hydrodynamic-to-collisionless crossover for
a variety of tractable models. Here we will present some additional
results which shed light on the role duality can play on the form of
these functions.

In Section \ref{sec:cp1} we will consider the role of duality in
Abelian systems, by examining the self-dual non-compact, easy-plane,
$\mathbb{CP}^1$ field theory discussed by Motrunich and
Vishwanath~\cite{mv}. Closely related results apply to other Abelian
CFTs whose particle-vortex duals have been described in the
literature \cite{witten,intrili,strassler1,strassler2,balents}, some
of which are supersymmetric (in which case, particle-vortex
duality is known as `mirror symmetry'). The Lagrangian formulation of
the
${\mathbb C}{\mathbb P}^1$ theory involves two complex scalar fields
and one gauge field $A_\mu$, which is coupled to a gauge current
$J_{1\mu}$. The theory has a global U(1)$\times Z_2$ symmetry, and
we will denote by $J_{2\mu}$ the Noether current arising from the
U(1) global symmetry. There is another conserved current, the
topological current $J_{\rm
top}^\mu=\epsilon^{\mu\nu\lambda}\partial_\nu A_\lambda$, which is
conserved by the Bianchi identity. The topological and Noether
currents exchange under the self-duality. As we will see in
Section~\ref{sec:cp1}, the two-point correlator of $J_{\rm top}^\mu$
is the inverse of that of $J_{1\mu}$. We use the notations of
Eqs.~(\ref{j0}), (\ref{j1}) with $a,b=1,2$.

The $Z_2$ symmetry ensures that the cross-correlations of the
$J_{1\mu}$, $J_{2\mu}$ currents vanish, and consequently there are
only two constants $K_{1}\equiv K_{11}$ and $K_{2}\equiv K_{22}$ in
Eq.~(\ref{j0}), and similarly for the $T>0$ functions in
Eq.~(\ref{j1}). We examine the duality transformations of these
function in Section~\ref{sec:cp1} and show that the existence
of a self-dual critical point leads to the functional relations%
\footnote{
   We only keep the one-photon irreducible (1PI) part
   in $K_1^{L,T}$, as explained in Section \ref{sec:cp1}.
}
\begin{subequations}
\label{cp1dual}
\begin{eqnarray}
K_1^L (\omega, k)\; K_2^T (\omega, k) &=& \frac{1}{\pi^2}\,,\\
K_2^L (\omega, k)\; K_1^T (\omega, k) &=& \frac{1}{\pi^2}\,,
\end{eqnarray}
\end{subequations}
which hold for general $T$, while for the constants in Eq.~(\ref{j0})
this implies $K_1 K_2 = 1/\pi^2$. Note that these relations are not
sufficient to determine the conductivities $\sigma_{1,2}(\omega/T)$;
from Eq.~(\ref{sigmak}), only their product obeys
$\sigma_1(\omega/T)\,\sigma_2(\omega/T ) = 1/\pi^2$, at all
$\omega/T$. Thus we expect that for this self-dual model, the
conductivities will remain non-trivial functions of $\omega/T$
exhibiting the hydrodynamic-collisionless crossover, and their
functional form has to be determined from the solution of a quantum
Boltzmann equation.

In Section~\ref{sec:m}, we turn to a field theory with non-Abelian
symmetries: the supersymmetric Yang Mills (SYM) gauge theory with a
SU($N$) gauge group and $\mathcal{N}{=}8$ supersymmetry
\cite{Seiberg}.
At long distances, the theory flows under the renormalization group
to a strongly coupled 2+1 dimensional $\mathcal{N}{=}8$
superconformal field theory (SCFT), which is believed to describe
degrees of freedom on a stack of $N$ M2-branes \cite{Sethi-Susskind,IMSY}.
In the
limit of large $N$, the SCFT can be analyzed by using the AdS/CFT
correspondence \cite{MAGOO}. The gravity description of the SCFT is
given by
M-theory on 3+1 dimensional anti-de Sitter space times a seven-sphere,
and in the large $N$ limit corresponds to 10+1 dimensional
supergravity on $\mathrm{AdS}_4\times S^7$. The AdS/CFT
correspondence provides a
method to compute real-time response functions at finite temperature
\cite{recipe,Herzog:2002pc}, in which case the gravity theory
contains a black hole in AdS$_4$. In the limit of low frequency and
momentum $\omega\ll T$, $k\ll T$ one finds hydrodynamic behavior
in the SCFT \cite{CH}.%
\footnote{
    Hydrodynamic charge transport at small $\omega$ and $k$
    is of course not specific to the $\mathcal{N}{=}8$ SCFT
    in 2+1 dimensions.
    Hydrodynamics from the supergravity description
    was first found in strongly coupled $\mathcal{N}{=}4$ SYM
    in 3+1 dimensions \cite{PSS-hydro}, and later in a variety
    of other strongly coupled field theories
   \cite{membrane,Buchel:2004hw,Benincasa:2006ei,Dp-Dq}.
    In strongly coupled ${\cal N}{=}4$ SYM in $D{=}3{+}1$,
hydrodynamic to
    collisionless crossover functions $K^{L,T}(\omega,k)$
    were computed in \cite{photons-sym}. Note that in $D=3+1$ the
conductivity
    is not dimensionless \cite{ssbook}, but is proportional to $T$ in
the hydrodynamic
    limit $\omega \ll T$.
}
The surprising solvability in this limit therefore demands our
attention.%
\footnote{
    Of course, there are other
    well-known $D=2{+}1$ CFTs which are solvable in the large $N$ limit,
    such as the $O(N)$ $\varphi^4$ field theory. However, all of these
    are theories of particles which are infinitely long-lived at
    $N=\infty$, and so do not exhibit hydrodynamic behavior in this
limit. Indeed,
    an infinite-order resummation of the $1/N$ expansion is
    invariably necessary \cite{ssbook} (via the quantum Boltzmann
    equation) to obtain hydrodynamics. These solvable
    theories become weakly coupled as $N\to\infty$, while the
    $\mathcal{N}{=}8$ SYM remains strongly coupled even as $N\to\infty$.
}

The 2+1 dimensional SCFT has a global SO(8) R-symmetry (the
symmetry of the seven-sphere in the supergravity description), and
therefore has a set of conserved currents $J^a_{\mu}$, $a=1,\ldots,
28$. The SO(8) symmetry implies that $K_{ab} = K \delta_{ab}$, and
so there is only a single universal constant $K$ at zero
temperature. Similarly, in Eq.~(\ref{j1}) there are only two
independent functions $K^L (\omega, k)$ and $K^T (\omega, k)$ which
characterize the CFT response at finite temperature. In
Section~\ref{sec:m} we will compute these functions in the
$N{\to}\infty$ limit, for all values of $\omega/T$ and $k/T$. We
also prove that these functions obey the identity
\begin{equation}
   K^L(\omega, k)\; K^T(\omega, k) = \frac{N^3}{18 \pi^2},
\label{mdual}
\end{equation}
at general $T$, which is strikingly similar to Eqs.~(\ref{cp1dual}).
Now this relation and Eq.~(\ref{sigmak}) do indeed determine $\sigma
(\omega/T)$ (and $K$) to be the frequency-independent constant which
is the square root of the right-hand-side of Eq.~(\ref{mdual}). In
other words, for this model, the hydrodynamic and high-frequency
collisionless conductivities are equal to each other. Nevertheless,
the theory does have a hydrodynamic-to-collisionless crossover at
all nonzero $k$ (as we will review in Section~\ref{sec:m}), where
$K^L(\omega,k)\neq K^T(\omega,k)$, and so Eq.~(\ref{mdual}) is not
sufficient to fix the correlators at $k{\neq}0$. Thus the identity
Eq.~(\ref{mdual}) causes all signals of the
hydrodynamic-collisionless crossover to disappear {\em only\/} at
$k{=}0$.

The similarity of Eq.~(\ref{mdual}) to Eq.~(\ref{cp1dual})
suggests that explanation of the frequency independence of the
conductivity of
the $\mathcal{N}=8$ SYM SCFT lies in a self-duality property.
Section~\ref{sec:m} demonstrates that this is indeed the case.
Under the AdS/CFT correspondence, the two-point
correlation function of the SO(8) R-currents in $D=2{+}1$ is
holographically
equivalent to the correlator of a SO(8) gauge field on an
asymptotically AdS$_4$ background. In the large $N$ limit, the
action of the SO(8) gauge field is Gaussian, and is easily shown to
possess electromagnetic (EM) self-duality under which the
electric and magnetic fields are interchanged. We demonstrate in
Section~\ref{sec:emduality} that it is precisely this EM self-duality
of the
3+1 dimensional gauge field
which leads to the constraint (\ref{mdual}) in the SCFT.
Thus the SYM theory obeys a self-duality which is not readily
detected in 2+1 dimensions,
but becomes explicit in the holographic theory in 3+1 dimensions.
The generalization of the particle-vortex duality of Abelian
CFTs in $D=2+1$ to non-Abelian CFTs is facilitated by the holographic
extension
to the theory on AdS$_4$.

There have been a few earlier studies connecting dualities in $D=4$
to those in $D=3$. Sethi \cite{sethi} considered the Kaluza-Klein
reduction
of S-duality from $D=4$ to $D=3$ by compactifying the $D=4$ theory
on a circle in one dimension. This is quite different from the
connection above,
using a holographic extension.
The work of
Witten \cite{witten} makes a connection which is the same as ours
above (see also
the work of Leigh and Petkou \cite{petkou1}).
He examined the connection between Abelian
particle-vortex duality (`mirror symmetry') of CFTs in $D=2+1$
to the action of SL(2,$Z$) on Abelian gauge theories
on AdS$_4$ at zero temperature.
We have considered a similar connection at non-zero temperature for the
${\cal N}{=}8$ SCFT, and shown that it is ``holographically self
dual'' in the large $N$ limit;
combined with
the non-Abelian SO(8) symmetry (which implies a single $K$), the
constraints
for the current correlators are stronger than those for Abelian
theories.

We will also consider in Appendix~\ref{app:d2}
other non-Abelian theories with known gravity descriptions.
In particular, we will show that for a theory on a stack of D2
branes, a non-trivial dilaton profile prevents EM self-duality.
In this case, we do not have the constraint (\ref{mdual}), and so find
a frequency dependent conductivity.

\section{Abelian, non-compact ${\mathbb C}{\mathbb P}^1$ model}
\label{sec:cp1}
\noindent
This section will consider duality properties and current
correlations of the Abelian, easy-plane ${\mathbb C}{\mathbb P}^1$
model of Ref.~\cite{mv}. This is a theory of two complex scalars
$z_{1,2}$ and a non-compact U(1) gauge field $A_\mu$; the
non-compactness is necessary to suppress instantons (monopoles), and
we indicate below Eq.~(\ref{j2w}) the modifications required when
monopoles are present.

More generally, one can consider dualities of the
non-compact ${\mathbb C}{\mathbb P}^{N-1}$ model
where the global SU($N$) flavor symmetry has been explicitly
broken down to U(1)$^{N-1} \times G_N$,
with $G_N$ some subgroup of the permutation group of $N$ objects \cite
{balents}.
The $N=1$ case, which is better known as
the Abelian Higgs model, will be described in Appendix~\ref{app:cs}. The
$N=2$ case (with $G_2 = Z_2$) is described below.
The $T>0$ results below have a generalization to all $N > 2$, with
the mappings spelled out in Ref.~\cite{balents}. Only the $N=2$
case is self-dual, and this is our reason for focusing on it.

It is interesting to note that the duality properties of the
non-compact ${\mathbb C}{\mathbb P}^{N-1}$ models have strikingly
similar
counterparts in $D=2+1$ theories with $\mathcal{N}=4$
supersymmetry \cite{intrili,strassler1,strassler2}.
In particular, the correspondence is to the theories with one
U(1) vector (gauge) multiplet and $N$ matter hypermultiplets (SQED-$N$).
SQED-1 is dual to a theory
of a single hypermultiplet, with no vector multiplet%
\footnote{The theory of a single
hypermultiplet is free. This is because the Gaussian fixed point is
protected by $\mathcal{N}=4$ supersymmetry \cite{strassler2}. In the
non-supersymmetric
case, the Gaussian fixed point is unstable to the interacting Wilson-
Fisher fixed point.};
this corresponds
to the duality, reviewed in Appendix~\ref{app:cs},
of the Abelian Higgs model to the theory of a single complex scalar
with no gauge field
(also known as the XY model or the O(2) $\varphi^4$ field theory).
Next, SQED-2 is self-dual, as is our $N=2$ case.
  For $N>2$, the dual of SQED-$N$ is a quiver gauge
theory, as is the case for the ${\mathbb C}{\mathbb P}^{N-1}$
models \cite{balents}.\footnote{%
  A quiver gauge theory consists of a direct product of gauge group
  factors along with matter fields transforming in the bifundamental
  representation of pairs of group factors.  The word quiver is used
  because the bifundamental fields
  are often represented as arrows.
}
Our results below for $T>0$
should have straightforward extensions to these $\mathcal{N}=4$
supersymmetric
theories.

\subsection{Conserved currents}

\begin{figure}
    \includegraphics[width=5in]{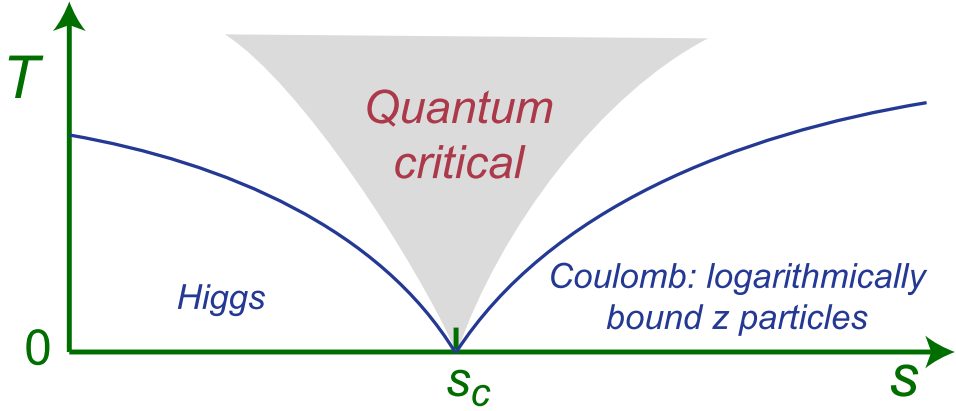}
\caption{Phase diagram \cite{mv} of the easy-plane non-compact ${\mathbb
C}{\mathbb P}^1$ model (Eq.~(\ref{sz})) in 2 spatial dimensions as a
function of the coupling $s$ and temperature $T$. The quantum
critical point is at $s=s_c$, $T=0$. The finite $T$ correlations of
the CFT describe the shaded quantum critical region; the boundary of the
shaded region is a crossover into a different physical region, not
a phase transition. The full lines
are Kosterlitz-Thouless (KT) phase transitions. The
KT line for $s<s_c$ describes the disappearance of quasi-long-range
$xy$ order
of $\vec{N}$. The KT transition for $s>s_c$
describes the deconfinement of $z$ quanta which are logarithmically
bound by the Coulomb interaction
in the low temperature phase into particle-anti-particle
pairs. The phase diagram can also be described in terms of the dual
$w$ theory in Eq.~(\ref{sw}). Duality interchanges the
two sides of $s=s_c$ ($T$ remains invariant under duality), and the $z
$ Coulomb phase
is interpreted as a $w$ Higgs phase and vice versa.} \label{phasediag}
\end{figure}

Let us now begin our analysis of the non-supersymmetric $N=2$ case.
The action of the non-compact ${\mathbb C}{\mathbb P}^1$ theory is
\begin{eqnarray}
\mathcal{S} &=& \int\!\! d^2\!x\, dt\; \Bigl[
\left|\left(\partial_\mu - i A_\mu\right) z_1 \right|^2 +
\left|\left(\partial_\mu - i A_\mu\right) z_2 \right|^2 + s \left(
|z_1|^2 + |z_2 |^2 \right) + u
\left(|z_1|^2 + |z_2|^2 \right)^2 \nonumber \\
&~&~~~~~~~~~~~~~~+ v |z_1|^2 |z_2 |^2 + \frac{1}{2e^2} \left(
\epsilon^{\mu\nu\lambda} \partial_\nu A_\lambda \right)^2 \Bigr],
\label{sz}
\end{eqnarray}
with $u>0$ and $-4u<v<0$. For these negative values of $v$, the
phase for $s$ sufficiently negative has $|\langle z_1 \rangle | =
|\langle z_2 \rangle | \neq 0$. We can also define a gauge-invariant
vector order parameter $ \vec{N} = z^\ast \vec{\sigma} z$, where
$\vec{\sigma}$ are the Pauli matrices, and the constraint $v<0$
implies that $\vec{N}$ prefers to lie in the $xy$ plane: hence
`easy-plane' (for $v>0$, $\vec{N}$ would be oriented along the $z$
`easy-axis', realizing an Ising order parameter). The ${\mathbb
C}{\mathbb P}^1$ model is usually defined with fixed length
constraint $|z_1|^2 + |z_2|^2 = 1$, but here we have only
implemented a soft constraint by the quartic term proportional to
$u$; we expect that the models with soft and hard constraints have
the same critical properties. We are interested in the nature of the
quantum phase transition accessed by tuning the value of $s$ to a
critical value $s=s_c$. For $s>s_c$, we have a `Coulomb' phase
$\langle \vec{N} \rangle = 0$ with a gapless photon, while for
$s<s_c$ there is a `Higgs' phase with $\langle \vec{N} \rangle \neq
0$. The phase diagram \cite{mv} of the model in the $s$, $T$ plane
is shown in Fig.~\ref{phasediag}.
Both the Higgs and Coulomb phases have phase transitions as the
temperature is raised: for the former it is driven by the loss of
the Higgs (quasi)-long-range order, while for the latter it is a
``confinement-deconfinement'' transition of the $z$
particle-anti-particle pairs formed from the logarithmic Coulomb
force. Neither of these transitions is of interest to us in this
paper. Rather, we will compute $T>0$ correlations of the CFT
associated with the quantum critical point, and these describe the
physical properties of the shaded quantum critical
region in Fig.~\ref{phasediag}.

The theory has a discrete $Z_2$ symmetry which exchanges $z_1$ and
$z_2$. The continuous symmetries are a gauge U(1) symmetry
\begin{equation}
z_1 \rightarrow z_1 e^{i \phi}~~;~~ z_2 \rightarrow z_2 e^{i
\phi}~~;~~A_\mu \rightarrow A_\mu + \partial_\mu \phi
\end{equation}
and a global U(1) symmetry
\begin{equation}
z_1 \rightarrow z_1 e^{i \varphi}~~;~~ z_2 \rightarrow z_2 e^{-i
\varphi}.
\end{equation}
Associated with these symmetries we can define two currents
\begin{equation}
J_{1\mu} = i \left( z_1^\ast (\partial_\mu - i A_\mu) z_1 - z_1
(\partial_\mu + i A_\mu) z_1^\ast \right) + i \left( z_2^\ast
(\partial_\mu - i A_\mu) z_2 - z_2 (\partial_\mu + i A_\mu) z_2^\ast
\right).
\end{equation}
and
\begin{equation}
J_{2\mu} = i \left( z_1^\ast \partial_\mu z_1 - z_1 \partial_\mu
z_1^\ast \right) - i \left( z_2^\ast \partial_\mu z_2 - z_2
\partial_\mu z_2^\ast \right) \ .
\end{equation}
Note that $J_1$ is even under the $Z_2$ symmetry, while $J_2$ is
odd. Current conservation implies that at $T>0$ these have two-point
correlators of the form in Eq.~(\ref{j1}), with the 4 distinct
functions $K_{1,2}^{L,T}$.

Now consider the correlators of the gauge field $A_\mu$. It is
useful to write this in terms of the leading quadratic terms in the
Coleman-Weinberg effective potential:
\begin{eqnarray}
W &=& \frac{1}{2} \int_{k,\omega} \Biggl\{ -(k_i A_0 + \omega A_i
)^2 \left[ \frac{1}{e^2} +
\frac{\Pi^L (k, \omega)}{-\omega^2 + k^2} \right] \nonumber \\
&~&~~+  A_i A_j \left(
  \delta_{ij} - \frac{k_i k_j}{k^2} \right)\left[ \frac{k^2}{e^2} +
\Pi^T (k, \omega) + \frac{\Pi^L (k, \omega) \omega^2 }{-\omega^2 +
k^2} \right] \Biggr\} + \ldots
\end{eqnarray}
where $\Pi^{L,T}$ are the two components of the photon self energy
(the `polarization' operator); these are related to the current
correlations by $\Pi^{L,T} = \sqrt{p^2} K^{L,T}_1$.

A key point is that at the conformal fixed point describing the
phase transition at the quantum critical point $s=s_c$ we can safely
take the limit $e \rightarrow \infty$ in the above. This is because
$\mbox{dim}[\Pi] = 1$, and so the induced polarizations are more
singular than the bare Maxwell term. This is a very generic property
of CFTs with gauge fields in $D=2{+}1$. From the effective potential
we can obtain the form of the gauge-invariant two-point correlators
in the critical regime (it is easiest to work this out in the Coulomb
gauge $k_i A_i = 0$):
\begin{eqnarray}
\left\langle \epsilon_{ij} k_i A_j ~;~ \epsilon_{i'j'} k_{i'} A_{j'}
\right\rangle &=& \frac{k^2}{\Pi^T (k, \omega)}\,, \nonumber
\\
\left\langle \epsilon_{i'j'} k_{i'} A_{j'} ~;~ (k_i A_0 + \omega
A_i) \right\rangle &=&  \epsilon_{i'i} \frac{\omega k_{i'}}{\Pi^T
(k,
\omega)}\,, \nonumber \\
\left\langle (k_i A_0 + \omega A_i)~;~ (k_j A_0 + \omega A_j)
\right\rangle &=& \left( \delta_{ij} - \frac{k_i k_j}{k^2} \right)
\frac{\omega^2}{\Pi^T (k, \omega)} - \frac{k_i k_j}{k^2}
\frac{(-\omega^2+k^2)}{\Pi^L (k, \omega)}\,. \label{aa}
\end{eqnarray}

\subsection{Vortices and duality}
\label{sec:dual}

Here we will build a dual description of the ${\mathbb
C}{\mathbb P}^1$ model, treating the vortices of
the original model as complex scalar fields in the dual
description.
Consider the topological vortex excitations in the Higgs state of
the action (\ref{sz}). These are characterized \cite{babaev} by a
pair of winding
numbers $(n_1, n_2)$ associated with the phases of $z_1$ and $z_2$
out at spatial infinity. In general, such a vortex has a
logarithmically diverging energy because the currents are only
partially screened by the gauge field $A_\mu$. By an extension of the
Abrikosov-Nielsen-Olesen argument, it can be seen that
the co-efficient of the logarithmically divergent energy is
proportional to
\begin{equation}
   \left(2 \pi n_1 - \int\!d^2x\,\epsilon_{ij} \partial_i A_j\right)^2
   +
   \left(2 \pi n_2 - \int\!d^2x\,\epsilon_{ij} \partial_i A_j\right)^2,
\end{equation}
and this is minimized when the total $A_\mu$ flux is quantized as
\cite{mv,babaev,balents}
\begin{equation}
   \int\!d^2x\,\epsilon_{ij} \partial_i A_j = \pi (n_1 + n_2).
\label{eq:flux}
\end{equation}
Let us now identify the $(1,0)$ vortex as the worldline of a dual
particle $w_1$, the $(0,1)$ vortex as the worldline of a dual
particle $w_2$, and try to construct a dual theory by
introducing complex scalar fields $w_1(x)$, $w_2(x)$.
Then from Eq.~(\ref{eq:flux}),
Lorentz covariance implies that the total $w$
current is related to the $A_\mu$ flux:
\begin{equation}
\frac{1}{\pi}\epsilon_{\mu \nu \lambda} \partial^\nu A^\lambda = i
\left( w_1^\ast \partial_\mu w_1 - w_1 \partial_\mu w_1^\ast \right)
+ i \left( w_2^\ast \partial_\mu w_2 - w_2
\partial_\mu w_2^\ast \right). \label{aw}
\end{equation}
A second key property is that there are forces with a logarithmic
potential between the $w_{1,2}$ particles. These are also easily
seen from the structure of the classical vortex solutions of
Eq.~(\ref{sz}). Also, it is the {\em difference\/} of the $z_1$ and
$z_2$ currents, which is not screened by the $A_\mu$ field, which
contributes to an {\em attractive \/} logarithmic potential between
the $w_{1}$ and $w_2$ particles. Another way to see this is to
consider the configuration of the gauge-invariant Higgs field $(N_x,
N_y)$ around each vortex: the $w_1$ has an anti-clockwise winding of
the $\mbox{arg}(N_x+iN_y)$, while the $w_2$ has a clockwise winding.
Because there is a finite stiffness associated with this Higgs
order, a $w_1$ particle will attract a $w_2$ particle, while two $w_1$
(or $w_2$) particles will repel each other.

We can now guess the form of the effective theory for the $w_{1,2}$
particles. We mediate that logarithmic potential as the Coulomb
potential due to a new `dual' gauge field $\widetilde{A}_\mu$. Then
general symmetry arguments and the constraints above imply the dual
theory \cite{mv}
\begin{eqnarray}
\widetilde{\mathcal{S}} &=& \int\!d^2x\, dt \Bigl[
\left|\left(\partial_\mu {-} i \widetilde{A}_\mu\right) w_1
\right|^2 + \left|\left(\partial_\mu {+} i \widetilde{A}_\mu\right)
w_2 \right|^2 + \widetilde{s} \left( |w_1|^2 + |w_2 |^2 \right) +
\widetilde{u}
\left(|w_1|^2 + |w_2|^2 \right)^2 \nonumber \\
&~&~~~~~~~~~~~~~~+ \widetilde{v} |w_1|^2 |w_2 |^2 +
\frac{1}{2\widetilde{e}^2} \left( \epsilon^{\mu\nu\lambda}
\partial_\nu \widetilde{A}_\lambda \right)^2 \Bigr]. \label{sw}
\end{eqnarray}
Note especially the difference in the charge assignments from
(\ref{sz})---now the $w_{1,2}$ particles have opposite charges under
$\widetilde{A}_\mu$. Apart from this, the theories have an identical
form, and so current correlation functions
$\widetilde{K}^{L,T}_{1,2}$, associated with the global and gauge
U(1) symmetries, will have the same dependence upon the couplings in
$\widetilde{\mathcal{S}}$ as the $K^{L,T}_{1,2}$ have on
$\mathcal{S}$. However, the explicit expressions for the current in
terms of the field operators have a sign interchanged:
\begin{equation}
\widetilde{J}_{1\mu} =
   i \left( w_1^\ast (\partial_\mu - i \widetilde{A}_\mu) w_1 - w_1
   (\partial_\mu + i \widetilde{A}_\mu)
    w_1^\ast \right)
  - i \left( w_2^\ast (\partial_\mu + i \widetilde{A}_\mu) w_2 - w_2
   (\partial_\mu - i \widetilde{A}_\mu) w_2^\ast
\right)
\end{equation}
and
\begin{equation}
\widetilde{J}_{2\mu} = i \left( w_1^\ast \partial_\mu w_1 - w_1
\partial_\mu w_1^\ast \right) + i \left( w_2^\ast \partial_\mu w_2 -
w_2
\partial_\mu w_2^\ast \right).
\label{j2w}
\end{equation}

We note in passing the extension of the above analysis to a {\em
compact\/} ${\mathbb C}{\mathbb P}^1$ theory of the $z$ particles.
Following Polyakov \cite{polyakov}, we have to include monopoles
which change the $A_\mu$ flux by $2\pi$. This can be achieved by
adding the term $-y_m (w_1 w_2 + w_1^\ast w_2^\ast)$ to the $w$
action $\widetilde{\mathcal{S}}$, where $y_m$ is the monopole
fugacity. This monopole operator is neutral under
$\widetilde{A}_\mu$ charge and, from Eq.~(\ref{aw}), catalyzes the
required change in $A_\mu$ flux. This is a relevant perturbation:
the theories for the $z$ and $w$ particles are no longer equivalent
under
duality,
and the universality class of the transition is changed. We will not
consider the compact case further; for more details, see the review
\cite{ssmott}.

Returning to the non-compact theory, we note the duality mapping can
now also be carried backwards from the $w$ theory to the $z$ theory,
and from (\ref{aw}) we see that the theories $\mathcal{S}$ and
$\widetilde{\mathcal{S}}$ are connected by the relations
\begin{subequations}
\label{dualj}
\begin{eqnarray}
\frac{1}{\pi}\epsilon_{\mu \nu \lambda} \partial^\nu A^\lambda &=&
\widetilde{J}_{2\mu}, \\
\frac{1}{\pi}\epsilon_{\mu \nu \lambda} \partial^\nu
\widetilde{A}^\lambda &=& J_{2\mu}.
\end{eqnarray}
\end{subequations}
  From these relations, Eq.~(\ref{aa}), and the definition (\ref{j1}),
we immediately obtain the relation between $K_1$ and $K_2$:
\begin{subequations}
\label{eq:KKdual}
\begin{eqnarray}
    &&  K_1^T(\omega,k)\, \widetilde K_2^L(\omega,k) = \frac{1}{\pi^2}
\,,\quad
    \widetilde K_1^T(\omega,k)\, K_2^L(\omega,k) = \frac{1}{\pi^2}\,,
    \\
    &&  K_1^L(\omega,k)\, \widetilde K_2^T(\omega,k) = \frac{1}{\pi^2}
\,,\quad
    \widetilde K_1^L(\omega,k)\, K_2^T(\omega,k) = \frac{1}{\pi^2}\,.
\end{eqnarray}
\end{subequations}
Now, assuming a single second-order transition obtained by tuning
the parameter $s$, the above reasoning implies that this critical
point must be self-dual, $K_1^{T,L}=\widetilde K_1^{T,L}$, and
$K_2^{T,L}=\widetilde K_2^{T,L}$. Self-duality thus immediately
implies relation (\ref{cp1dual}), as claimed in the Introduction.

Monte Carlo simulations \cite{proko} of a current loop model related
to $\mathcal{S}$ observe a weak first-order transition. This is
possibly because they are using a particular lattice action which is
not within the domain of attraction of the self-dual point. In any
case, the duality mappings between the two phases on either side of
the transition apply, and the constraints on a possible CFT remain
instructive.

\section{The M2-brane theory}
\label{sec:m}

This section examines the transport properties of the non-Abelian
SU($N$) Yang Mills
theory in $D=2{+}1$ with $\mathcal{N}=8$ supersymmetry. The weak-
coupling action and
field content
of this theory is most directly understood by dimensional reduction
of the $\mathcal{N}=1$ SYM theory in $D=9{+}1$ on the flat torus $T^7
$ \cite{tasi}.
This reduction shows that the $D=2{+}1$ theory has an explicit SO(7)
R-charge global symmetry. The $D=9{+}1$ SYM theory has only a single
gauge
coupling constant, and therefore, so does the $D=2{+}1$ theory. The
latter
coupling has a positive scaling dimension, and flows to strong-coupling
in the infrared. It is believed \cite{Seiberg} that the flow
is to an infrared-stable fixed point that describes a SCFT.
It was also argued that this SCFT has
an emergent R-charge symmetry which is expanded to SO(8). We shall
be interested in the transport properties of this SO(8) R-charge
in the SCFT at $T>0$ in the present section.

We are faced by a strongly-coupled SCFT, and a perturbative analysis
of the field theory described above is not very useful. Instead,
remarkable progress is possible using the connection to string theory
and the AdS/CFT correspondence. The $D=2{+}1$ SYM theory is contained
in the low energy description of Type IIA string theory in the presence
of a stack of $N$ D2-branes. The flow to strong coupling of the SYM
theory corresponds in string theory to the lift of ten-dimensional
Type IIA strings to eleven-dimensional M-theory \cite{MAGOO}.
So we can directly access the $D=2{+}1$ SYM SCFT by considering M-theory
in the presence of a stack of $N$ M2-branes \cite{IMSY}. In the large
$N$ limit, M-theory can be described by the semiclassical
theory of eleven-dimensional supergravity, and this
will be our main tool in the analysis described below. This formulation
also makes the SO(8) R-charge symmetry explicit, because the M2-branes
curve the spacetime of
eleven-dimensional supergravity to AdS$_4 \times S^7$.

Another powerful feature of the supergravity formulation is that
it can be extended to $T>0$. We have to consider supergravity in a
spacetime
which is asymptotically AdS$_4$, but which also contains a black hole.
The Hawking temperature of the black hole then corresponds to the
temperature of the SCFT \cite{wittenm} (for example,
fluctuation-dissipation theorems are satisfied~\cite{Herzog:2002pc}).
Hydrodynamics of the SCFT
emerges from the semiclassical supergravity dynamics in the presence
of the
black hole.\footnote{%
  Strictly speaking, the appearance of a black hole is dual to being
at finite
  temperature \emph{and} being in a deconfined phase; it is possible
to have a finite temperature
  gravitational description without a black hole \cite{wittenm,
HerzogPRL}.
}

Turning to our explicit computation of dynamics in M-theory, we
consider
the gravitational background associated with
a stack of $N$ M2-branes, with $N \gg 1$ \cite{andy,IMSY,CH},
\begin{equation}\label{metric}
   ds^2 = \frac{r^4}{R^4} \left[ -f(r) dt^2 + dx^2 + dy^2\right]
         + \frac{R^2}{r^2}
         \left[ \frac{dr^2}{f(r)} + r^2d\Omega_7^2 \right],
\end{equation}
where $f(r)=1-r_0^6/r^6$. It is more convenient for us to change
coordinates from $r$ to $u=(r_0/r)^2$, in terms of which
\begin{equation}
   ds^2 = \frac{r_0^4}{R^4u^2}[-f(u)dt^2+dx^2+dy^2]
   + \frac{R^2}{4u^2f}du^2 + R^2 d\Omega_7^2
\end{equation}
and $f(u)=1-u^3$. The horizon of the black hole is located at $u=1$, and
the boundary of AdS$_4$ is at $u=0$.

The relationship between the quantities in the worldvolume SCFT
($N$ and temperature $T$) and those of the metric ($R$ and $r_0$)
are given by~\cite{IMSY,CH}
\begin{equation}
   \pi^ 5 R^9 = \sqrt 2\, N^{3/2}\kappa^2, \qquad
   T = \frac3{2\pi} \frac{r_0^2}{R^3}\,,
   \label{kelevennorm}
\end{equation}
where $\kappa$ is the gravitational coupling strength of
$D=10{+}1$ supergravity.

There is a precise correspondence between correlation functions
computed in the D=2{+}1 CFT and correlation functions of
supergravity fields computed in the metric
(\ref{metric}) \cite{MAGOO,recipe,Herzog:2002pc}.
We will use this to compute charge transport properties.

In the metric~(\ref{metric}) a 7-sphere factors out:
$R^2d\Omega_7^2$. The spacetime thus has a SO(8) symmetry. This
matches with the global symmetry in the M2 worldvolume theory: there
is a R-charge which transforms under the same global symmetry. The
following subsections will compute the two-point correlations of the
R-charge currents, $J_{a\mu}$, with $a=1, \ldots, 28$.

The existence of a compact 7-sphere makes it possible to do
Kaluza-Klein reduction on this space. We expand all fields in terms
of spherical harmonics on the 7-sphere. The original fields of
M-theory are the metric tensor $g_{\mu\nu}$ and a three-index
antisymmetric tensor $A_{\mu\nu\lambda}$. Upon Kaluza-Klein
reduction, an SO(8) gauge field
appears from the components of the
metric and the three-form where only one index is in the AdS$_4$
directions ($t$, $x$, $y$, and $u$) and the others are in the $S^7$
directions (see Appendix~\ref{app:g} for details).
The action for this gauge field is
\begin{equation}\label{M2action}
   S = -\frac 1{4\gfourD^2}
       \int\!d^4x\,\sqrt{-g}\, g^{MA} g^{NB}
       F^a_{MN} F^a_{AB},
\end{equation}
where uppercase Latin indices $A$, $B$, $M$, $N$ run four values of
$t$, $x$, $y$, and $u$ (in contrast to Greek indices $\alpha$,
$\beta$, $\mu$, $\nu$ which run $t$, $x$ and $y$). The
four-dimensional gauge coupling constant $\gfourD$ is {\em
dimensionless}, and its large $N$ value is computed in
Appendix~\ref{app:g}
\begin{equation}
   \frac1{\gfourD^2} = \frac{\sqrt{2}}{6 \pi} N^{3/2}.
\end{equation}

Although we focus on the gravity background constructed from a stack
of $N$ M2-branes in flat 11-dimensional space, there are a number
of related examples which are
easily understood from considering (\ref{M2action}).
The key observation, which we discuss further in Section~\ref
{sec:emduality},
is that (\ref{M2action}) exhibits
classical electric-magnetic duality.
In the case of our M2-brane theory, this duality is close enough to a
self-duality to
enforce a relation on the current-current two point functions and result
in a frequency independent conductivity.
In fact, this self-duality holds in a more general context.
Consider
an eleven dimensional space which factorizes into $\mathbb{R}^{2,1}$
and a Calabi-Yau four-fold which develops a local singularity.  By
placing
a stack of M2-branes at the singularity, we should obtain a more exotic
2+1 dimensional conformal field theory which still has at least a U(1)
  global R-symmetry.  Kaluza-Klein reduction of the gravity theory
will yield
  precisely (\ref{M2action}) and our results on holographic self-duality
  will carry over to these more general cases.

  There are two other interesting generalizations to consider in which
  holographic self-duality fails.
  After Kaluza-Klein reduction, the gauge fields $F_{AB}$ will support
electrically charged black holes \cite{duffliu}.  These
black holes are dual to introducing an R-charge chemical potential to
the field
theory.  
Another
interesting 2+1 dimensional field theory with a holographic
description is the theory living on a stack
of D2-branes in type IIA string theory.
In both cases, there is generically a nontrivial scalar
which
appears in a modification of (\ref
{M2action}) as a coupling
constant which depends on the holographic radial direction.  The
relation on the two-point
functions will be between a theory with coupling $\gfourD(u)$ and one
with coupling $1/\gfourD(u)$.
For details concerning this more general perspective, see Appendix
\ref{app:d2}.

\subsection{Current-current correlators}
\noindent
We now proceed to the computation of the two-point correlators of
the $J_{a\mu}$ in the CFT at $T>0$. Here we will work in Minkowski
space (real frequencies and time), and so define the current
correlation as follows:
\begin{equation}
     C_{\mu\nu}(x-y) \delta_{ab} =
     -i\,\theta(x^0{-}y^0) \langle [J_{a\mu}(x),J_{a\nu}(y)]\rangle.
     \label{defC}
\end{equation}
The $\delta_{ab}$ follows from SO(8) symmetry. The expectation value
is taken in a translation-invariant state, so we can Fourier
transform to $C_{\mu\nu}(p)$, where $p_\mu=(-\omega,\k)$.
Spectral density is proportional to the imaginary part of the
retarded function,
\begin{equation}
   \rho_{\mu\nu}(p)=-2\Im C_{\mu\nu}(p).
\end{equation}
It is an odd, real function of $p$, whose diagonal components are
positive (for positive frequency).
Expectation values of all global conserved charges are assumed to
vanish in the equilibrium state; in other words we consider systems
without chemical potentials.
Conservation of $J_{a\mu}(x)$ implies that the correlation functions
may be defined so that they satisfy the Ward identity\footnote{%
     One may choose to define the correlation functions in such a way
that
     local (in position space) counter-terms appear on the right-hand
side
     of the Ward identities.
     The correlation functions defined in this way will differ from
     $C_{\mu\nu}(p)$ by analytic functions
     of $\omega$ and $\k$.}
$p^\mu C_{\mu\nu}(p) = 0$.
Then, as in Section~\ref{sec:intro} and in Eq.~(\ref{j1}), we can
write $C_{\mu \nu}$ in the form
\begin{equation}
    C_{\mu\nu}(p) = P_{\mu\nu}^T\, \Pi^T(\omega,k) +
                    P_{\mu\nu}^L\, \Pi^L(\omega,k)\ .
\label{eq:C-rotation-inv}
\end{equation}
(The relationship between $\Pi$ and $K$ is $\Pi^{T,L}= \sqrt{p^2} K^
{T,L}$.)
Without loss of generality one can take the spatial momentum
oriented along the $x$ direction, so that $p=(\omega,k,0)$. Then the
components of the retarded current-current correlation function are
\begin{equation}
    C_{yy}(\omega,k) = \Pi^T(\omega,k)\ ,
\end{equation}
as well as
\begin{equation}
    C_{tt}{=} \frac{k^2}{\omega^2{-}k^2}\, \Pi^L(\omega,k),\ \
    C_{tx}{=}C_{xt}{=}\frac{-\omega k}{\omega^2{-}k^2}\, \Pi^L
(\omega,k),\ \
    C_{xx}{=} \frac{\omega^2}{\omega^2{-}k^2}\, \Pi^L(\omega,k)\ .
      \label{eq:Czz}
\end{equation}

\subsection{Correlation functions from AdS/CFT}
\label{sec:ads-cft}
\noindent
In order to find the retarded function, one needs to study
fluctuations of vector fields on the background spacetime created by
a stack of M2-branes. At the linear order the fields satisfy the
equations
\begin{equation}
   \d_M (\sqrt{-g}\, g^{MA}g^{NB} F_{AB}) =0.
\end{equation}
These equations are to be solved with the boundary conditions
\begin{equation}
   \lim_{u\to 0} A_\mu (u, x) = A_\mu^0(x),
\end{equation}
at $u{=}0$. Near $u{=}1$ one imposes the outgoing-wave boundary
condition, which means that for $u$ slightly less than 1 the
solution is purely a wave that propagates toward the horizon. Due to
translational invariance with respect to $x$ one can solve for each
Fourier mode $e^{ip\cdot x}$ separately. The result can be
represented in the form
\begin{equation}\label{AFA0}
   A_\mu(u, p) = {M_\mu}^\nu(u,p) A_\nu^0(p).
\end{equation}
Then, according to the AdS/CFT prescription formulated in
Ref.~\cite{recipe}, the current-current correlator can be found from
the formula%
\footnote{
    Greek indices on $M_{\mu\nu}$ are raised using the flat space
    Minkowski metric.
}
\begin{equation}\label{C-prescr}
   C_{\mu\nu}(p) = -\chi \lim_{u\to0} M_{\mu\nu}'(u,p),
\end{equation}
where $\chi$ is the constant that appears in the normalization of
the action,
\begin{equation}
   S= \frac{\chi}{2}\int \!du\, d^3 x\left(A_t^{\prime2}
      - f A_x^{\prime2} - f A_y^{\prime2} +\dots \right)\,,
\end{equation}
(only terms with two derivatives with respect to $u$ are written).
In our case $\chi={4\pi T}/{3\gfourD^2}$. It turns out that $\chi$
is precisely the charge susceptibility.\footnote{%
     The hydrodynamic density-density response function found in
     \cite{CH} is $ C_{tt} = (1/\gfourD^2){k^2}/(i\omega{-}D_c k^2)$.
     Comparing this to the hydrodynamic form $C_{tt} = \chi D_c k^2/(i
     \omega{-}D_c k^2)$, we find
     the above value for charge susceptibility $\chi$.
}

The prescription given above might appear ad-hoc. However
it is a special case of a more general AdS/CFT prescription that
gives real-time correlators of any number of
operators~\cite{Herzog:2002pc}. For our task, however, the above
prescription is technically most straightforward to implement.

We work in the radial gauge $A_u=0$, and take all fields $A_\mu(x)$
to be proportional to $e^{-i\omega t+i\k\cdot{\bm x}}$. Taking
momentum $\k$ along the $x$ direction, $\k=(k,0)$, one finds that
the fluctuating vector fields satisfy the following equations
\cite{CH}
\begin{eqnarray}
   \wn A_t'+\qn f A_x' =0\,, && \label{eq:AtAx}\\
   A_t''-\frac{1}{f}(\wn\qn A_x + \qn^2 A_t)=0\,, &&
   \label{eq:Atpp}\\
   A_x'' + \frac{f'}{f} A_x' + \frac{1}{f^2} (\wn\qn A_t + \wn^2 A_x)
=0\,, &&
      \label{eq:Axpp}\\
   A_y'' + \frac{f'}{f} A_y' + \frac{1}{f^2} (\wn^2-\qn^2 f)A_y =0\,.&&
   \label{eq:Ay}
\end{eqnarray}
Here prime denotes derivative with respect to $u$; $\wn$ and $\qn$
are the dimensionless frequency and momentum, $\wn\equiv
3\omega/(4\pi T)$, $\qn\equiv 3k/(4\pi T)$. Note that the equation
for the transverse potential $A_y$ decouples from the rest.
Moreover, Eq.~(\ref{eq:Axpp}) can be shown to follow from
Eqs.~(\ref{eq:AtAx}) and (\ref{eq:Atpp}) and so is not independent.
Combining Eqs.~(\ref{eq:AtAx}) and (\ref{eq:Atpp}) one can obtain an
equation that does not involve $A_x$,
\begin{equation}\label{eq:Atppp}
   A_t''' + \frac{f'}f A_t'' + \frac1{f^2}(\wn^2-\qn^2f) A_t' =0.
\end{equation}
One can think about this equation as a second-order equation for
$A_t'$. It was observed in \cite{CH} that {\it Eq.~(\ref{eq:Atppp})
has the same form as the equation for $A_y$}. Such degeneracy is
unusual, and we now proceed to explore its implications.

\subsubsection{Transverse channel}
\noindent
Let us start with the retarded function for transverse currents,
$C_{yy}(\omega,\k)$. According to the AdS/CFT
prescription~(\ref{C-prescr}),
\begin{equation}
   C_{yy}(p) = -\chi \lim_{u\to0} M_{yy}'(u,p).
\end{equation}
The function $M_{yy}(u,p)$ is the solution to Eq.~(\ref{eq:Ay})
which satisfies the outgoing-wave boundary condition on the horizon
$u{=}1$, and $M_{yy}(0,p)=1$ at the boundary $u{=}0$.

Let us denote a solution to Eq.~(\ref{eq:Ay}) which satisfies the
outgoing boundary condition at the horizon as $\psi(u)$. The
normalization of $\psi(u)$ is left arbitrary. Near $u{=}0$,
Eq.~(\ref{eq:Ay}) allows two asymptotic solutions, which can be
expressed in terms of the Frobenius series,
\begin{eqnarray}
   Z_I(u) &=& 1+h Z_{II}(u) \ln u + b_{I}^{(1)}u +\dots,\\
   Z_{II}(u) &=& u(1+b_{II}^{(1)}u + b_{II}^{(2)}u^2+\dots).
\end{eqnarray}
The coefficient $b_{I}^{(1)}$ is arbitrary, and we set it to zero.
All other coefficients are determined by substituting expansion
(\ref{eq:AB}) in the original equation (\ref{eq:Ay}). In particular,
we find that $h{=}0$, therefore
\begin{eqnarray}
   Z_I(0) &=& 1, \qquad Z_I'(0) = 0,\nonumber\\
   Z_{II}(0) &=& 0, \qquad Z_{II}'(0) = 1.\label{ZZp}
\end{eqnarray}
The outgoing-wave solution $\psi(u)$ can be exprressed as
\begin{equation}
   \psi(u) = \CA Z_{I}(u) + \CB Z_{II}(u),
\label{eq:AB}
\end{equation}
where $\CA$ and $\CB$ depend on the parameters of the equation, in
particular on $\wn$ and $\qn$. From Eq.~(\ref{ZZp}) it follows that
$\psi(0)=\CA$ and $\psi'(0)=\CB$. The properly normalized mode
function is $M_{yy}(u,p)=\psi(u)/\psi(0)$, and therefore we find
\begin{equation}
   C_{yy}(\wn,\qn) = - \chi\, \frac{\CB(\wn,\qn)}{\CA(\wn,\qn)}\,.
\end{equation}

\subsubsection{Longitudinal channel}
\noindent
Let us now look at the correlators in the longitudinal channel:
$C_{tt}$, $C_{tx}$, and $C_{xx}$. For that we need to solve
Eqs.~(\ref{eq:AtAx}) and (\ref{eq:Atpp}).
First, we know that $A_t'(u)$ satisfies the same equation as
$A_y(u)$. Therefore, we can write $A_t'(u)= c\psi(u)$, where $c$ is
some coefficient. This coefficient can be fixed from the boundary
conditions at $u=0$ by employing Eqs.~(\ref{eq:Atpp}) and
$\psi'(0)=\CB$. We find
\begin{equation}
   A_t'(u) = \left[ \frac\CA\CB Z_I(u) + Z_{II}(u)\right]
             (\wn\qn A_x^0 + \qn^2 A_t^0).
\end{equation}
  From Eq.~(\ref{eq:AtAx}) we also find
\begin{equation}
   A_x'(u) = -\frac1f \left[ \frac\CA\CB Z_I(u) + Z_{II}(u)\right]
             (\wn^2 A_x^0 + \wn\qn A_t^0).
\end{equation}
These equations are to be compared with Eq.~(\ref{AFA0}), from which
one extracts $M_{\mu\nu}'(u,p)$. Putting $u=0$, one find the
correlators
\begin{equation}
   C_{tt}(\wn,\qn) = \chi \qn^2\, \frac{\CA(\wn,\qn)}{\CB(\wn,\qn)}\,,
   \quad\quad
   C_{xx}(\wn,\qn) = \chi \wn^2\, \frac{\CA(\wn,\qn)}{\CB(\wn,\qn)}\,.
\end{equation}
In Appendix \ref{app:soln} we show that at zero momentum, $\qn{=}0$,
the mode equation (\ref{eq:Ay}) can be solved analytically,
which allows one to determine $\Pi^T(\wn,0)=\Pi^L(\wn,0)$.
However, one can determine the conductivity without
explicitly solving the mode equation, as we now show.

\subsection{Conductivity}
\label{sec:conductivity}
\noindent
We see that both $C_{yy}$ and $C_{xx}$ are expressed in terms of the
same connection coefficients $\CA$ and $\CB$. Eliminating the
coefficients, we find
\begin{equation}
   C_{xx}(\wn,\qn)C_{yy}(\wn,\qn) = -\chi^2 \wn^2\,,\quad\quad
   C_{tt}(\wn,\qn)C_{yy}(\wn,\qn) = -\chi^2 \qn^2\,.
\label{eq:cc}
\end{equation}
Expressed in terms of the self-energies $\Pi^T$, $\Pi^L$ this reads
\begin{equation}
   \Pi^T(\wn,\qn)\, \Pi^L(\wn,\qn) = -\chi^2 (\wn^2{-}\qn^2).
\label{eq:pipi}
\end{equation}
Note that this relation holds for all $\wn$ and $\qn$: we have not
made any small-frequency approximations anywhere. In fact, we did
not even have to solve the mode equations! Combining
Eqs.~(\ref{j1}), (\ref{eq:C-rotation-inv}), and (\ref{eq:pipi}), we
obtain our main result in Eq.~(\ref{mdual}).

As discussed in Section~\ref{sec:intro}, at zero momentum, rotation
invariance implies that $\Pi^T{=}\Pi^L$, therefore relation
(\ref{eq:pipi}) uniquely determines the self-energy%
\footnote{
     Up to a sign, which can be fixed by requiring positivity
     of the spectral function $\rho_{yy}=-2\,\Im \Pi^T$.
}
$\Pi^T(\omega,0)=\Pi^L(\omega,0)=-i\chi\wn$ for all $\wn$. The
conductivity is given by $\sigma(\omega/T)=i\Pi^T(\omega,0)/\omega$,
and we find
\begin{equation}
   \sigma(\omega/T)=\chi \frac{3}{4\pi T} = \chi D_c = \frac{1}
{\gfourD^2},
\label{eq:sigma}
\end{equation}
where $D_c=3/(4\pi T)$ is the diffusion constant found in \cite{CH}.
Note that the Einstein relation between the conductivity and the
diffusion
constant is satisfied. Also, as noted earlier, it is surprising that
$\sigma(\omega/T)$ is actually independent of $\omega/T$.
[Dependence
upon $\omega/T$ is found at all non-zero $k$, as is shown below.]
This $\omega$-independence is a consequence of the relation
(\ref{eq:pipi}), which in turn follows from the fact that $A_t'$ and
$A_y$ satisfy the same equation in the bulk.
It can be traced back to the electromagnetic duality of the
classical action (\ref{M2action}), as we now show.

\subsection{Electric-magnetic duality}
\label{sec:emduality}
\noindent
Even though the origin of the relation (\ref{eq:pipi})
is puzzling from the point of view of the microscopic
degrees of freedom in the ${\cal N}{=}8$ SCFT,
its origin from the bulk point of view
can be traced to electric-magnetic (EM) duality
of an abelian gauge field.
Indeed, current-current correlators are computed from the
Maxwell equations in the four-dimensional bulk, and it is
precisely in four dimensions that Maxwell equations may possess
EM duality.

Although in general the R-symmetry may be non-abelian and hence
be dual to a non-abelian gauge field in the bulk, we work in the
classical
supergravity limit and must keep $N$ large.
At large $N$, the gauge coupling $g_{4\rm{D}} \propto N^{-3/4}$ is
very small, and our non-abelian gauge field factorizes into a number of
effectively abelian pieces to leading order in $1/N$.

If we write equations of motion in terms of the gauge-invariant
$F_{MN}$ (rather than the vector potential), then Maxwell equations
have to be supplemented by a Bianchi identity,
\begin{subequations}
\label{eq:ME}
\begin{eqnarray}
   && \partial_M(\sqrt{-g} \, F^{MN}) = 0 \\
   && \partial_M(\sqrt{-g} \,
      \frac12 \eps^{MNAB} F_{AB}) =0\,,
\end{eqnarray}
\end{subequations}
where $\eps^{MNAB}$ is the totally antisymmetric tensor,
with $\eps^{0123}=1/\sqrt{-g}$.
Now, one can introduce $G_{MN}$ defined as
$F^{MN}=\frac12\eps^{MNAB}G_{AB}$,
which can be inverted to give
$G^{MN}=-\frac12 \eps^{MNAB}F_{AB}$.
Expressed in terms of $G$, the equations of motion become
\begin{subequations}
\label{eq:ME-dual}
\begin{eqnarray}
   && \partial_M(\sqrt{-g} \,
      \frac12 \eps^{MNAB} G_{AB}) =0\,,\\
   && \partial_M(\sqrt{-g} \, G^{MN}) =0\,.
\end{eqnarray}
\end{subequations}
Maxwell equations for $F$ become a Bianchi identity for $G$,
and vice versa.
$G_{MN}$ is the dual field strength tensor, and we can also
define a dual vector potential $B_M$ by
$G_{MN}=\partial_M B_N-\partial_N B_M$.
Note that the validity of EM duality does not depend
on the background spacetime having any particular symmetries
such as Lorentz symmetry, or rotational symmetry.

  From the point of view of AdS/CFT, the EM dual theory in the bulk
will correspond to some theory on the boundary, which is a dual
of the original SCFT.
In particular, the dual vector potential $B_\mu$ will couple to the
dual current $\tilde J_\mu$, and one can compute two-point functions
$C_{\mu\nu}^{\rm dual}(\omega,k)$ in the dual theory.

In components we have $F^{tz}=G_{xy}/\sqrt{-g}$.
This means that the equation for $\sqrt{-g}F^{tz}$
obtained from equations (\ref{eq:ME}) is the same as the equation
for $G_{xy}$, obtained from the dual equations (\ref{eq:ME-dual}).
In our particular example of the non-extremal M2 background metric,
we have $\sqrt{-g}F^{tu}\propto A_t'(u)$, and $G_{xy}\propto kB_y(u)$
(in the radial gauge).
Thus the equation for $A_t'(u)$ is the same as the equation for $B_y
(u)$.
Then, by the argument in section \ref{sec:ads-cft} we
find a relation between the self-energies $\Pi^{T,L}$
in the original theory, and the self-energies $\widetilde\Pi^{T,L}$
in the dual theory:
\begin{subequations}
\begin{eqnarray}
  && \Pi^T(\wn,\qn)\, \widetilde\Pi^L(\wn,\qn) = -\chi^2 (\wn^2{-}
\qn^2)\,,\\
  && \widetilde\Pi^T(\wn,\qn)\, \Pi^L(\wn,\qn) = -\chi^2 (\wn^2{-}
\qn^2) \ .
\end{eqnarray}
\end{subequations}
  For our M2-branes, EM duality is a self-duality, and the EM dual
theory
is the same as the original theory, as is evident
from equations (\ref{eq:ME}), (\ref{eq:ME-dual}).
Therefore, $C_{\mu\nu}=C_{\mu\nu}^{\rm dual}$,
and $\widetilde\Pi^T=\Pi^T$, $\widetilde\Pi^L=\Pi^L$.
This gives back our main result (\ref{eq:pipi}).%
\footnote{
     The present discussion assumes that the coupling constant
     $\gfourD^2$ is not inverted
     in the dual theory, which is justified for a free,
     sourceless, abelian gauge field.
     One could formally repeat the same steps leading to
     Eq.~(\ref{eq:pipi}), assuming $\widetilde{g}_{\rm 4D}^2=1/
\gfourD^2$,
     as is standard in EM duality.
     However, in this case the coupling constant
     $\widetilde{g}_{\rm 4D}^2\propto N^{3/2}$ becomes large,
     invalidating the bulk description in terms of a classical
     gauge field.
}
In the case when there are non-trivial background profiles
for scalar fields, the EM dual theory is not equivalent
to the original theory.
This is discussed in Appendix \ref{app:d2}.

\subsection{Full spectral functions}
\begin{figure}
   \begin{picture}(0,0)(0,0)
   \put(110,-10){$\wn$}
   \put(320,-10){$\wn$}
   \end{picture}
   \includegraphics[width=2.8in]{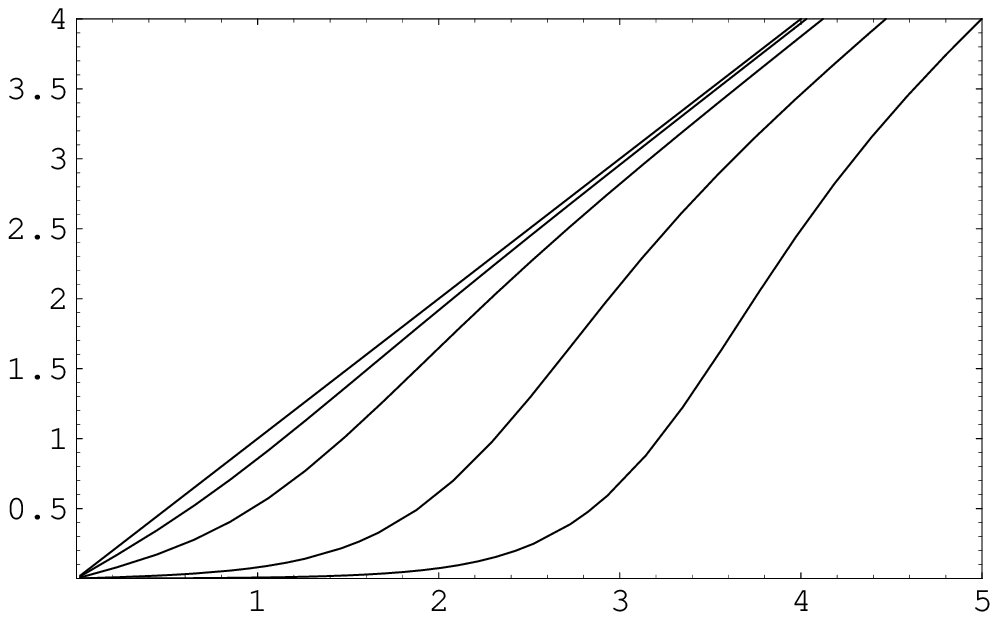}
   \includegraphics[width=2.8in]{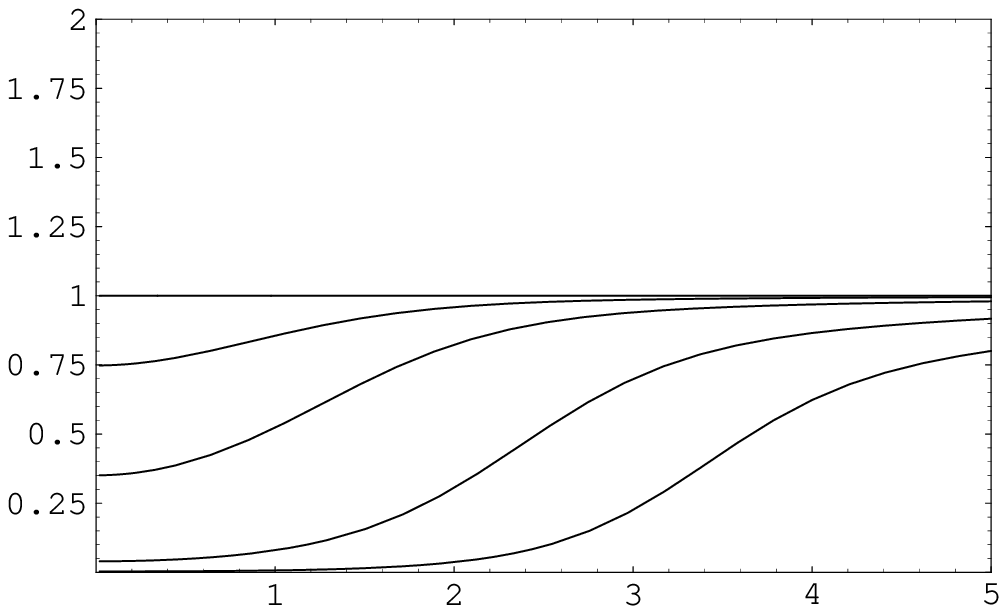}
\caption{
   Imaginary part of the retarded function $C_{yy}(\omega,k)$, plotted
   in units of $(-\chi)$,
   as a function of dimensionless frequency $\wn\equiv 3\omega/(4\pi
T)$,
   for several values of dimensionless momentum $\qn\equiv 3k/(4\pi T)$.
   Curves from left to right correspond to $\qn=0,0.5,1.0,2.0,3.0$.
   Left: $\Im C_{yy}(\wn,\qn)$,
   Right: $\Im C_{yy}(\wn,\qn)/\wn$.
} \label{fig:ImCyy}
\end{figure}
\begin{figure}
   \begin{picture}(0,0)(0,0)
   \put(110,-10){$\wn$}
   \put(320,-10){$\wn$}
   \end{picture}
   \includegraphics[width=2.8in]{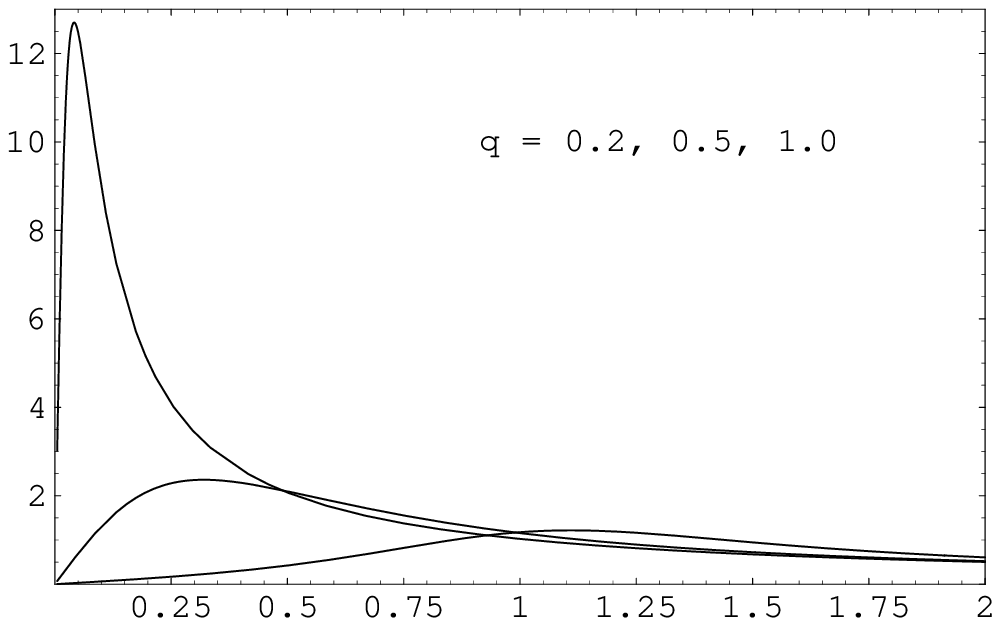}
   \includegraphics[width=2.8in]{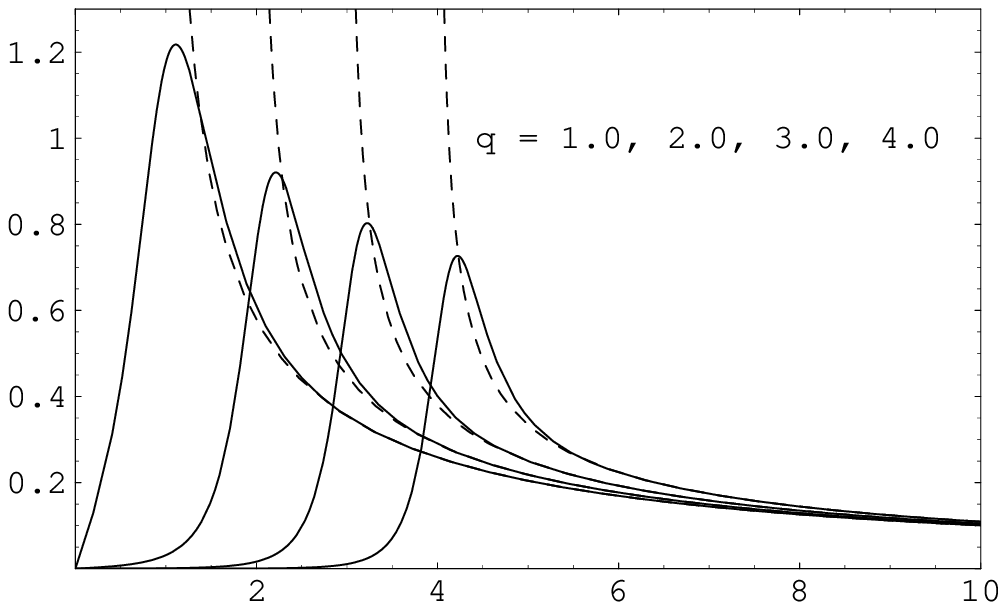}
\caption{
   Imaginary part of the retarded function $C_{tt}(\wn,\qn)/\qn^2$,
   plotted in units of $(-\chi)$,
   as a function of dimensionless frequency $\wn\equiv 3\omega/(4\pi
T)$,
   for several values of dimensionless momentum $\qn\equiv 3k/(4\pi T)$.
   Curves from left to right correspond to $\qn=0.2,0.5,1.0$
   (left panel), and $\qn=1.0,2.0,3.0,4.0$ (right panel).
   The dashed curves are plots of Eq.~(\ref{chigh}) divided by $k^2$.
} \label{fig:ImCtt}
\end{figure}
\noindent
We will now evaluate the spectral functions numerically, for all
$\omega$ and $k$. To do so, we find a solution $\psi(u)$ to the mode
equation (\ref{eq:Ay}) with the outgoing boundary conditions at the
horizon $u{=}1$. Then, as described in Section \ref{sec:ads-cft},
the retarded two-point function $C_{yy}(\omega,k)$ is proportional
to $\psi'(0)/\psi(0)$, while $C_{tt}(\omega,k)$ is proportional to
$\psi(0)/\psi'(0)$.

Figure \ref{fig:ImCyy} shows the imaginary part of the transverse
current-current correlation function, plotted in units of $(-\chi)$.
At zero momentum, $\Im C_{yy}$ is a linear function of
$\wn \equiv 3 \omega/(4\pi T)$ for
all $\wn$, as shown in the previous subsection. At large frequency,
the spectral function asymptotes to $\Im C_{yy}\sim(-\chi)\wn$,
regardless of the value of $\qn \equiv 3k/(4 \pi T)$.

The longitudinal correlators are directly related to the conserved
R-charge density, and so are more direct probes of hydrodynamic
behavior, and the hydrodynamic-to-collisionless crossover. Figure
\ref{fig:ImCtt} shows the imaginary part of the density-density
correlation function divided by $\qn^2$. At small momentum and
frequency, one clearly sees the diffusive peak, consistent with the
hydrodynamic expression in Eq.~(\ref{j0d})
\begin{equation}
\Im C_{tt}(\omega,k) = D_c\chi\frac{-\omega k^2}{\omega^2+(D_c
k^2)^2}~~~,~~~\mbox{$|\omega| \ll T$ and $k \ll T$.} \label{clow}
\end{equation}
At large frequency, the asymptotic form of the spectral function is
expected to be determined by the `collisionless' ground state
correlator. The latter was presented in Eq.~(\ref{j0n}), and here
has the form
\begin{equation}
\Im C_{tt}(\omega,k) = \frac{1}{\gfourD^2} \mbox{sgn}(\omega)
\frac{(-k^2)}{\sqrt{\omega^2 - k^2}}~~~,~~~|\omega|-k \gg T.
\label{chigh}
\end{equation}
Fig.~\ref{fig:ImCtt}, right, shows that this form is indeed well
obeyed. Indeed, Eqs.~(\ref{clow}) and (\ref{chigh}) are exactly the
correlators expected across a hydrodynamic-to-collisionless
crossover in a generic system \cite{forster}: the prefactor of $k^2$
in Eq.~(\ref{chigh}) is required by charge conservation even at
large $\omega$, while the factor of $1/\sqrt{\omega^2 - k^2}$ is set
by the CFT current scaling dimension and Lorentz invariance.

\begin{figure}
   \begin{picture}(0,0)(0,0)
   \put(100,-10){$\qn$}
   \end{picture}
\includegraphics[width=2.5in]{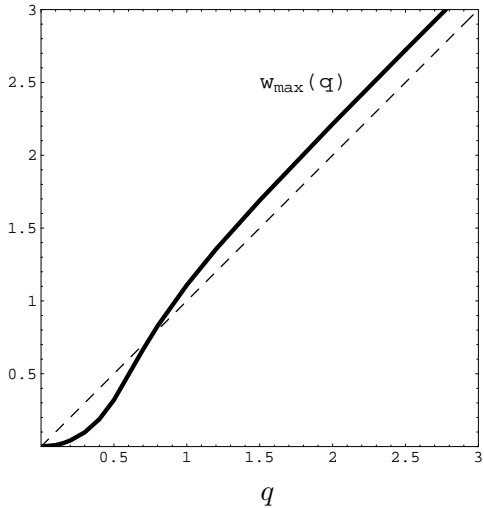}
\caption{The position of the peak of the spectral function in Fig.~
\ref{fig:ImCtt}.
          The dashed line is $\wn=\qn$.}
  \label{fig:wmaxq}
\end{figure}

In Fig.~\ref{fig:wmaxq}, we illustrate the crossover from the
hydrodynamic regime to the collisionless regime. For each value of
$\qn$ we find the value $\wn_{\rm max}$ where the function
$\Im C_{tt}(\wn, \qn)$ reaches its maximal value, and plot the resulting
function $\wn_{\rm max}(\qn)$. As we see on Fig.~\ref{fig:wmaxq}, at
small $\qn$ the location of the peak is $\wn_{\rm max}=\qn^2$, in
accordance with hydrodynamics. At large $\qn$ it slowly reaches the
asymptotic collisionless behavior $\wn_{\rm max}=\qn$.

What is unexpected, is that the two prefactors in Eqs.~(\ref{clow})
and (\ref{chigh}), $D_c \chi$ and $\gfourD^{-2}$, happen to be equal
to each other, as we saw in Eq.~(\ref{eq:sigma}). We have also seen
that this surprising feature is a consequence of the general
functional relations in Eqs.~(\ref{eq:pipi}) and (\ref{mdual}). As
we have discussed, such functional relations are not expected to
apply to a typical $D=2{+}1$ CFT, but only those which enjoy special
self-duality symmetries. Here the self-duality of the gauge
theory on AdS$_4$ led to the identical form
of Eqs.~(\ref{eq:Ay}) and (\ref{eq:Atppp}) which was shown eventually
to lead to Eqs.~(\ref{eq:pipi}) and (\ref{mdual}). In Appendix~\ref
{app:d2},
we consider a R-symmetry gauge field action with a non-trivial dilaton
which spoils the holographic self-duality and the frequency
independent conductivity. The field theory on a D2-brane
in type IIA string theory is an example with such a dilaton.

\section{Conclusions}
\label{sec:conc}

We considered finite temperature charge transport of quantum field
theories in
$D=2+1$ dimensions: the easy-plane $\mathbb{CP}^1$
model, and the CFT living on a stack of $N$ M2-branes in M-theory
(the $\mathcal{N}=8$, SU($N$) SYM theory).
In the former theory, Abelian particle-vortex
self-duality imposes a relationship (Eq.~(\ref{cp1dual})) between
different current correlators. In the latter theory, we found a
strikingly similar relationship (Eq.~(\ref{mdual}))
between longitudinal and transverse
components of the correlators of the SO(8) R-charge.
This relationship led to a frequency-independent conductivity
for the M2 worldvolume theory at zero wavevector, but hydrodynamic
behavior and the hydrodynamic-collisionless crossover did appear
at non-zero wavevectors. We also demonstrated that
for the D2-brane theory, our argument
for frequency independent conductivity fails because of a
nontrivial dilaton background.

We traced the origin of the SO(8) charge correlation constraint
of the SYM theory, and its frequency-independent conductivity,
to an electromagnetic self-duality of the holographic
theory on AdS$_4$. Thus, the generalization of
three-dimensional Abelian particle-vortex duality to non-Abelian
theories
becomes manifest only after a holographic extension to a four-
dimensional
theory. For Abelian theories, the AdS/CFT connection between particle-
vortex
duality in three dimensions and the SL(2,$Z$) invariance of four-
dimensional
Abelian gauge theories was explored earlier in \cite{witten,petkou1}.

Our results for the SU($N$) SYM theory were established at large $N$.
Does holographic self-duality,
and the relationship%
\footnote{Of course, the constant on the right-hand-side of Eq.~(\ref
{mdual}) would
have finite $N$ corrections. The issue is whether the right-hand-side
remains
independent of $\omega$ and $k$ for $T>0$ also at finite $N$.}
Eq.~(\ref{mdual}), hold also for finite $N$? The fact that the large
$N$ theory
has hydrodynamic behavior is evidence for
the ``generic'' nature of this limit. Furthermore, Eq.~(\ref{mdual})
has the same structure as Eq.~(\ref{cp1dual}), and the latter is
believed to be
an exact relationship, obtained without a large $N$ limit.
While these facts are encouraging, establishing self-duality
at finite $N$ requires looking at the full M-theory
on AdS$_4$. Its low energy limit is $\mathcal{N}=8$ supergravity
\cite{cremmer,dewit,duff,duffliu,freund}
(Section~\ref{sec:m} considered only the SO(8) gauge
fields of this theory), and its ``generalized E$_{7(7)}$ duality
invariance'' \cite{dewit} (which appears to include EM duality)
has remnants in M-theory \cite{hull}.

It would be very interesting
to find an Abelian field theory which obeyed a relationship
as simple as Eq.~(\ref{mdual}), found here for the SYM theory.
An unsuccessful attempt to find such a theory is
described in Appendix~\ref{app:cs}. The closest we could get
is Eq.~(\ref{cp1dual}), obeyed by the easy-plane
$\mathbb{CP}^1$ model \cite{mv} and its expected generalization to
the SQED-2 theory with $\mathcal{N}=4$
supersymmetry \cite{intrili,strassler1,strassler2}. A fundamental
feature
of Abelian particle-vortex duality is exchange of U(1)
`flavor' and `topological' currents, and we have not been able
to construct a theory in which these currents are equivalent to
each other (which would lead to a single $K$ in Eq.~(\ref{j0})).
However, non-Abelian theories can have additional symmetries which
rotate different U(1) currents into each other; this was important
for the simplicity of Eq.~(\ref{mdual}).

Finally, we would like to emphasize that the unexpected relation
between the self-energies found in this paper,
\begin{equation}
   K^L(\omega,k)\, K^T(\omega,k) = {\rm const}\ ,
\label{eq:mm}
\end{equation}
holds beyond the ${\cal N}=8$ SYM theory.%
\footnote{
    As described in Appendix \ref{app:g}, there is a whole class of
    2+1 dimensional CFTs satisfying Eq.~(\ref{eq:mm}).
    For large-$N$ field theories which are
    dual to M-theory on $AdS_4 \times X$, where
    $X$ is a seven dimensional Sasaki-Einstein manifold, with currents
    normalized as in Appendix~\ref{app:g}, the value of
    the constant in the right-hand side of Eq.~(\ref{eq:mm}) is
    ${N^3}/({2 \pi^{10}})\, \Vol(X)^2$.
}
It applies to the CFTs whose electromagnetic response is described by the
Maxwell action (\ref{M2action}) in the 3+1 dimensional
asymptotically AdS space. Thus the relation (\ref{eq:mm}) should be
viewed as another example of universality that characterizes
finite-temperature response in the AdS/CFT correspondence. Previous
examples of such universality include the universal value of the
viscosity to entropy density ratio $\eta/s=1/4\pi$ \cite{KSS}, and a
possible universal value of the friction coefficient for a heavy
particle \cite{CH-R}. Unlike these other examples, the universal
relation (\ref{eq:mm}) applies only to 2+1 dimensional CFTs at
finite temperature. On the other hand, unlike these other examples,
the universal relation (\ref{eq:mm}) applies at arbitrary $\omega$
and $k$.

\acknowledgments We thank M.~Ernebjerg, J.~Liu, D.~Shih, M.~Strassler,
A.~Strominger,
C.~Vafa, and A.~Vishwanath for useful
discussions. This research was supported by the NSF under grants
PHY99-07949 (PK and CH) and DMR-0537077 (SS), and by the DOE under
grants
DE-FG02-96ER40956 (CH) and DE-FG02-00ER41132 (DTS).
C.H. thanks the organizers of the
String Phenomenology workshop at the KITP, UCSB.
P.K. thanks the organizers of the INT workshop
``From RHIC to LHC: achievements and opportunities''
at the University of Washington,
and the Harvard University Physics Department, where part
of this work was completed.

\appendix

\section{Thermal correlators of CFT\lowercase{s} in $D=1{+}1$}
\label{app:cft2}

First, let us consider an arbitrary Lorentz-scalar observable
$\mathcal{O}$ of a CFT in $D=1{+}1$ with scaling dimension $h$.
Then, at $T=0$, its two-point correlator in Euclidean space is
\begin{equation}
\left. C_{\mathcal{O}} (\tau, x) \right|_{T=0} \sim \frac{1}{(x^2 +
\tau^2)^h},
\label{cft2a}
\end{equation}
while the corresponding correlator in momentum and imaginary
frequency is $\sim (\omega^2 + k^2)^{h-1}$. By the conformal map
from the infinite plane to the cylinder with circumference $1/T$, we
can obtain the form of the correlation at $T>0$:
\begin{equation}
C_{\mathcal{O}} (\tau,x) \sim \left[\frac{\pi^2 T^2}{\sin(\pi T (\tau
- i
x))\sin(\pi T (\tau + i
x))}\right]^h. \label{cft2b}
\end{equation}
Notice that this expression is periodic in $\tau$, with period
$1/T$. Now let us Fourier transform Eq.~(\ref{cft2b}) to momenta $k$
and Matsubara frequencies $\omega_n$; because of the periodicity,
the $\omega_n$ must be integer multiples of $2 \pi T$, and the result is
\begin{equation}
C_{\mathcal{O}} (i\omega_n, k) \sim T^{2h-2} \frac{\Gamma (1-h)}{\Gamma (h)} \frac{ \Gamma \left(
\displaystyle \frac{h}{2} + \frac{ |\omega_n|+ik}{4 \pi T} \right)
\Gamma \left(
\displaystyle\frac{h}{2} + \frac{ |\omega_n|-ik}{4 \pi T} \right)}
{\Gamma \left(
\displaystyle 1-\frac{h}{2} + \frac{ |\omega_n|+ik}{4 \pi T} \right)
\Gamma \left(
\displaystyle 1-\frac{h}{2} + \frac{ |\omega_n|-ik}{4 \pi T}
\right)} . \label{cft2c}
\end{equation}
Finally, we
analytically continue this expression to real frequencies
from the upper-half frequency plane ($\omega_n > 0$)
with the mapping $i \omega_n \rightarrow \omega$ to obtain the
retarded two-point correlator at $T>0$.
This is a non-trivial function of $\omega$ and $k$, which describes
relaxation of $\mathcal{O}$ correlations at $T>0$. See Chapter 4 of
Ref.~\cite{ssbook} for more details.

Now let us consider the special case of a conserved current, and
search for the collisionless-to-hydrodynamic crossover. In this
case, $h=1$. Note that Eq.~(\ref{cft2c}) has a pole at $h=1$ with a residue
which is $\omega$ and $k$ independent; this reflects a logarithmic cutoff
dependence in the Fourier transform of Eq.~(\ref{cft2b}), and the finite $\omega$
and $k$ dependent contribution is obtained by subtracting the pole.
However, the current is not a Lorentz
scalar, so the above results do not directly apply anyway.
The density correlator at $T=0$ in Euclidean space
is
\begin{equation}
\left. C_{tt} ( \tau,x) \right|_{T=0} \sim \frac{1}{(\tau - i x)^2} +
\frac{1}{(\tau + i
x)^2} . \label{cft2d}
\end{equation}
In momentum and real frequency space, the Fourier transform of this
is cutoff independent because of the non-zero Lorentz spin:
\begin{equation}
\left. C_{tt} (\omega, k) \right|_{T=0} \sim \frac{-k^2}{k^2 -
\omega^2}. \label{cft2e}
\end{equation}
This is, of course, the generalization of Eq.~(\ref{j0n}) to
$D=1{+}1$. We can obtain the $T>0$ density correlator by a conformal
mapping of Eq.~(\ref{cft2d}), as was done earlier in
Eq.~(\ref{cft2b}); here the corresponding expression is
\begin{equation}
C_{tt} ( \tau,x) \sim \left[\frac{\pi T}{\sin(\pi T (\tau - i
x))}\right]^2 + \left[\frac{\pi T}{\sin(\pi T (\tau + i
x))}\right]^2
\label{cft2f}.
\end{equation}
Finally, let us Fourier transform Eq.~(\ref{cft2f}) to momentum and
Matsubara frequency space. Carrying out this transformation yields
an initially surprising result. Although the real space result in
Eq.~(\ref{cft2f}) depends upon temperature, the
$T>0$ result in momentum and frequency space has the same form as
that at $T=0$ in Eq.~(\ref{cft2e}):
\begin{equation}
C_{tt} (i\omega_n, k) \sim \frac{-k^2}{k^2 + \omega_n^2}.
\label{cft2g}
\end{equation}
The inverse Fourier transforms of Eqs.~(\ref{cft2e}) and (\ref{cft2g}) differ
only because the frequency $\omega_n$ is discrete, while $\omega$ is continuous.
So there is no hydrodynamic behavior at $T>0$, and no analog of the
result in Eq.~(\ref{j0d}).

The physical interpretation of the absence of hydrodynamic behavior
is simple. CFTs in $D=1{+}1$ can be holomorphically factorized, and
consequently, there are no interactions or collisions between left
and right movers. To obtain collisions, one has to consider the
influence of formally irrelevant perturbations which can couple left
and right movers. Only then will hydrodynamic behavior emerge: see
Ref.~\cite{giamarchi}. In contrast, in $D=2{+}1$,
hydrodynamics emerges already in the conformal scaling limit
\cite{damle}.

\section{Abelian duality with one complex scalar}
\label{app:cs}

Here we will make some remarks on the duality properties of theories
of a single complex scalar coupled to a U(1) gauge field with a
Chern-Simons term in $D=2+1$. Such theories have been studied
extensively in the context of the quantum Hall effect
\cite{leefisher,fradkin,pryadko,shimshoni,burgess,witten,wen}. A $
\mathcal{N}=3$
supersymmetric generalization of the theory below has been studied by
Kapustin and Strassler \cite{strassler2} and their results are very
similar to our $T=0$ results below. We will also present results for
the theory without the Chern-Simons term (whose supersymmetric
analog, noted in Section~\ref{sec:cp1},
is the $\mathcal{N}=4$ SQED-1 theory \cite
{intrili,strassler1,strassler2}).

We consider a theory with the following action, which is essentially
the single scalar version of Eq.~(\ref{sz}), with an
additional Chern-Simons term:
\begin{eqnarray}
\mathcal{S}_{\rm cs} &=& \int\!\! d^2\!x\, dt\; \Biggl[
\left|\left(\partial_\mu - i A_\mu\right) z \right|^2 + s |z|^2  + u
|z|^4  \nonumber \\
&~&~~~~~~~~~~~~~~~~~~~~~~~~ + \frac{1}{2e^2} \left(
\epsilon^{\mu\nu\lambda} \partial_\nu A_\lambda \right)^2 + \frac{
\alpha}{4 \pi} \epsilon^{\mu\nu\lambda} A_\mu
\partial_\nu A_\lambda  \Biggr] \ .
\label{sza}
\end{eqnarray}
In general, this theory is not a CFT. However, as in
Section~\ref{sec:cp1}, we can imagine accessing a second-order phase
transition out of a Higgs phase
at a critical value of the ``mass'' term $s=s_c$; we are
interested here in the duality properties of such a CFT.

First, standard methods \cite{peskin,dasgupta} can be used to obtain
a dual version of the action $\mathcal{S}_{\rm cs}$: we can either
use the continuum arguments of Section~\ref{sec:dual}, or apply
Poisson summation methods to a lattice discretization
\cite{fradkin,pryadko,burgess}. From this we obtain a dual field
theory, which has the same formal structure at long wavelengths:
\begin{eqnarray}
\widetilde{\mathcal{S}}_{\rm cs} &=& \int\!\! d^2\!x\, dt\; \Biggl[
\left|\left(\partial_\mu - i \widetilde{A}_\mu\right) w \right|^2 +
\widetilde{s} |w|^2  + \widetilde{u}
|w|^4  \nonumber \\
&~&~~~~~~~~~~~~~~~~~~~~~~~~ + \frac{1}{2 \widetilde{e}^2} \left(
\epsilon^{\mu\nu\lambda} \partial_\nu \widetilde{A}_\lambda
\right)^2 + \frac{\widetilde{\alpha}}{4 \pi}
\epsilon^{\mu\nu\lambda} \widetilde{A}_\mu
\partial_\nu \widetilde{A}_\lambda  \Biggr] \ .
\label{szb}
\end{eqnarray}
The similarity between $\mathcal{S}_{\rm cs}$ and
$\widetilde{\mathcal{S}}_{\rm cs}$ is encouraging and suggests that
we may be able to use
it to define a self-dual CFT. However, we will now argue that this
is not the case.

In general, the relationship of the coupling constants in
$\mathcal{S}_{\rm cs}$ and $\widetilde{\mathcal{S}}_{\rm cs}$ is
non-universal, and dependent upon the nature of the ultraviolet
cutoff (with one exception, see below).
However, there are a number of crucial constraints, which
are readily apparent from the explicit transformations. From these
constraints we find that there are 2 distinct sets of theories which
are connected
by duality:\\
~\\
\underline{Class A: Theories with no Chern-Simons terms:} These
theories have $\alpha = \widetilde{\alpha} = 0$. Then the duality
mappings show that we must have either $e=0$ or $\widetilde{e} = 0$
but {\em not\/} both \cite{peskin,dasgupta}; this is because
in a theory with zero electric charge,
duality maps the
coefficient of the matter kinetic energy (the ``stiffness'') to the
electric charge squared
of the dual theory.
Without loss of generality, let us choose
$e=0$. Then we may set $A_\mu = 0$, which
then defines $\mathcal{S}_{\rm cs}$ as
the theory of a single scalar with a global U(1) symmetry
(the XY model). The theory $\widetilde{\mathcal{S}}_{\rm cs}$ is the
Abelian Higgs model which has a gauged U(1) `symmetry'. Thus we have
obtained the familiar duality \cite{dasgupta} between the XY
model and the Abelian Higgs model in $D=2+1$. It is evident from the
distinct nature
of these models that they are not self-dual \cite{peskin,dasgupta}.\\
~\\
\underline{Class B: Theories with Chern-Simons terms:} Now both
$\alpha$ and $\widetilde{\alpha}$ must be non-zero, and indeed the
lattice duality tranformations show that they satisfy
\begin{equation}
\alpha \widetilde{\alpha} = -1, \label{alpha}
\end{equation}
and this is the only relationship between the couplings of
$\mathcal{S}_{\rm cs}$ and $\widetilde{\mathcal{S}}_{\rm cs}$ which is
universal.
Furthermore, the requirement that either $e$ or $\widetilde{e}$
vanish no longer appears; in general, both are non-zero and finite.
The duality also shows that it is not possible to eliminate the
kinetic terms of both gauge fields {\em i.e.\/} it is not possible
to set both $e = \infty$ and $\widetilde{e} = \infty$. Even if {\em
e.g.\/} we eliminate the gauge kinetic term in $\mathcal{S}_{\rm
cs}$ by setting $e=\infty$, then the duality yields a
finite $\widetilde{e}$ because, by
the particle-vortex prescription, the kinetic energy of
$\widetilde{A}$ is related to the kinetic energy of the $z$
particles, and the latter is finite. Because we are searching for a
self-dual theory, we need to keep both $e$ and
$\widetilde{e}$ finite. The implication of a finite $e$ (or
$\widetilde{e}$) is that the flux-attachment transformation
associated with the Chern-Simons term is `smeared out': each $z$
particle world-line has a total of $2\pi/\alpha$ $A_\mu$ flux
attached, but this flux is spread out over a finite length scale
determined by $e$. This smearing also means that the transformation
$1/\alpha \rightarrow 1/\alpha + 1$ does not map the theory onto
itself. This transformation is the $T$ operation defined by Witten
\cite{witten}, who also found that $T$ did not leave the theory
invariant. On the other hand, Fradkin and Kivelson \cite{fradkin}
claimed $T$ invariance for their model, which was defined in terms of
infinitely-thin particle and flux world-lines on a lattice with long-
range
interactions. It is unclear to us whether their
model can be mapped to a local continuum
action for a CFT.

Let us now consider correlators of the field theories without a
Chern-Simons term, in class A. As discussed above, we choose the
theory $\mathcal{S}_{\rm cs}$ to have $e=0$ and $\alpha=0$, so
this describes the O(2) $\varphi^4$ theory (the XY model).
We are interested in
the CFT at some critical $s=s_c$. The two-point correlator of the
U(1) current, $C_{\mu\nu}$, of $\mathcal{S}_{\rm cs}$ obeys
Eqs.~(\ref{j0},\ref{j1}) with a single constant $K$, and a single
set of functions $K^{L,T} (\omega, k)$. Similarly, the dual theory,
$\widetilde{\mathcal{S}}_{\rm cs}$ (which has $\widetilde{e} \neq 0$,
$\widetilde{\alpha}=0$ and is the Abelian Higgs model),
has a correlator $\widetilde{C}_{\mu\nu}$, and the corresponding
$\widetilde{K}$. Then the analog of the duality considerations of
Section~\ref{sec:cp1} imply that
\begin{eqnarray}
K^T (\omega, k)\; \widetilde{K}^L (\omega, k) &=& \frac{1}{4 \pi^2} \ ,
\nonumber \\
K^L (\omega, k)\; \widetilde{K}^T (\omega, k) &=& \frac{1}{4 \pi^2}
\, , \label{sdual}
\end{eqnarray}
and its $T=0$ limit $K \widetilde{K} = 1/(4 \pi^2)$.
This theory in class A is not self-dual, so the above relations do
not allow us to determine the conductivities $\sigma (\omega/T)$ and
$\widetilde{\sigma} (\omega/T) $, and only constrain their product.

Next, we consider correlators of class B. The field theory
$\mathcal{S}_{\rm cs}$ defines a CFT at some $s=s_c$, and we ask if
this CFT can be self-dual. At $T=0$, we have to generalize the form
of the current correlator
$C_{\mu\nu}$ from Eq.~(\ref{j0}) to \cite{fradkin,burgess,witten}
\begin{equation}
\left. C_{\mu\nu} (p) \right|_{T=0} = K \sqrt{p^2}  \left( \eta_{\mu
\nu} - \frac{p_\mu
p_\nu}{p^2} \right) + H \epsilon_{\mu\nu\lambda} p^\lambda \,,
\label{ja}
\end{equation}
where $K$, $H$ are two real constants characterizing the CFT. Note that,
at the gapless conformal fixed point,
there is no simple relationship%
\footnote{The one-loop expression for $H$ obtained from $\mathcal{S}_
{\rm cs}$
is exact as long as $s \neq
s_c$,
but the CFT at $s=s_c$ has corrections at all orders \cite{ssqhe,wen}.}
between the coupling constant
$\alpha$ and the constant $H$, although a theory in class B is
expected to have a non-zero $H$. At $T>0$, the generalization of
Eq.~(\ref{j1}) is
\begin{equation}
   C_{\mu\nu}(\omega , \k)
    = \sqrt{p^2} \Bigl( P^T_{\mu\nu}\, K^T (\omega, k)
    + P^L_{\mu\nu}\, K^L (\omega, k) \Bigr) + H (\omega, k)
    \epsilon_{\mu\nu\lambda} p^\lambda \,,
\label{ja1}
\end{equation}
with 3 distinct functions of $\omega/T$ and $k/T$ on the right hand
side; note that even the Hall conductivity (equal to $H(\omega, 0)$)
is a function of $\omega/T$ \cite{ssqhe}.
Similarly, we can also consider the
dual-correlator $\widetilde{C}_{\mu\nu}$ of the theory (\ref{szb})
and define a corresponding set of parameters
$\widetilde{K}$ and $\widetilde{H}$. The analog
\cite{leefisher,fradkin,pryadko,burgess,witten} of the arguments in
Section~\ref{sec:cp1} shows the following exact relationship between
these parameters at $T=0$:
\begin{equation}
(H+iK) ( \widetilde{H} + i \widetilde{K}) = - \frac{1}{4 \pi^2} \,.
\label{sl2}
\end{equation}
The real and imaginary parts of Eq.~(\ref{sl2}) generalize the $T=0$
limit of Eq.~(\ref{sdual}) to class B. For $T>0$, we have
\begin{equation}
\widetilde{K}^T (\omega, k)  = \frac{K^T (\omega, k)}{D(\omega, k)}~~;~~
\widetilde{K}^L (\omega, k)  = \frac{K^L (\omega, k)}{D(\omega, k)}~~;~~
\widetilde{H} (\omega, k)  = -\frac{H (\omega, k)}{D(\omega, k)}\,,
\label{ja2}
\end{equation}
with
\begin{equation}
D(\omega, k) \equiv 4 \pi^2 \left( K^T (\omega, k) K^L (\omega, k) +
H^2 (\omega, k) \right) \,.
\end{equation}
Note that Eqs.~(\ref{ja2}) reduce to Eqs.~(\ref{sdual}) when $H=0$, and
to Eq.~(\ref{sl2}) at $T=0$.

For the class B model to be self-dual, we clearly need
$\widetilde{K}=K$ and $\widetilde{H} = H$. From Eq.~(\ref{sl2}) we
observe that this is only possible for $K=1/(2 \pi)$ and $H=0$.
However, a
model with $H=0$, which surely requires $\alpha=0$, is not in class
B. It is in class A, and we argued earlier that a class A model
could not be self-dual.

To conclude, although the model $\mathcal{S}_{\rm cs}$, and its dual
$\widetilde{\mathcal{S}}_{\rm cs}$, define interesting
CFTs, with their correlators obeying Eqs.~(\ref{sdual},\ref{sl2},\ref
{ja2}),
we have shown that such CFTs cannot be self-dual. This conclusion is in
accord with those of Kapustin and Strassler \cite{strassler2}
and Witten \cite{witten} on
related models.

\section{Normalization of gauge field action on A\lowercase{d}S$_4$}
\label{app:g}

\noindent
The R-symmetry gauge field can be thought of as arising from
Kaluza-Klein reduction of an 11 dimensional supergravity solution on
a regular positive curvature Sasaki-Einstein manifold $X$ of real
dimension seven.
The size of the gauge group is determined by the isometry group of $X$.
For instance, when $X= S^7$, the group is SO(8).  By definition,
Sasaki-Einstein
manifolds have at least one U(1) isometry.
In this section, we normalize the U(1) R-symmetry gauge field action
in terms of
the eleven dimensional gravitational coupling using results of
Ref.~\cite{BHK}.  Although the identification of this gauge field as
a combination of
metric and $F_4$ form perturbations in $D=11$ supergravity predates
Ref.~\cite{BHK}
(see \cite{DuffPopeWarner}),
Ref.~\cite{BHK} provides a convenient starting point for considering
issues of
normalization.   The normalization is not sensitive to temperature,
and hence it is
convenient to work here at $T=0$.

To first order, the vector potential $A$ perturbs the
eleven dimensional metric as follows:
\be
ds^2 = \frac{r^2}{L^2} \eta_{\alpha \beta} dx^\alpha dx^\beta + L^2
\frac{dr^2}{r^2}
+ 4 L^2 ds_X^2 \ ,
\ee
where
\be
ds_X^2 = \left( \frac{q}{4} \right)^2 \left( d\psi + \frac{4}{q}
\sigma + \frac{2}{q} A \right)^2
+ h_{a \bar b} dz^a d\bar z^b \ .
\ee
The Minkowski tensor $\eta_{\alpha\beta}$ runs over
the three coordinates $x^0$, $x^1$, and $x^2$.  Together the
coordinates $x^i$ and $r$
give four dimensional anti-de Sitter space with radius of curvature $L
$.\footnote{%
The relation to $R$ in the body of the paper is $2L = R$.
}
Here $h_{a \bar b}$ is a K\"ahler-Einstein metric on a complex three
dimensional manifold
we will call $V$.
Setting $A=0$, $X$ would be a U(1) fibration over the three-fold,
giving rise
to a real seven-dimensional Sasaki-Einstein manifold.  The one form $
\sigma$ is
constructed such that $d\sigma = 2 \omega$ where $\omega$ is the K
\"ahler form
on $V$.
With the angle $\psi$ constrained to lie between 0 and $2 \pi$,
the integer $q$ obeys the relation $\omega = \pi q c_1 / 4$ where $c_1
$ is
the first chern class of the U(1) fibration.  In general $q=1$, but
in certain
cases where $c_1(V)$ is divisible, $q$ may be more. For instance, in
the case of $S^7$,
$X$ is  a U(1) fibration over ${\mathbb{CP}^3}$ and $q=4$.
In \cite{BHK}, the relation between $\psi$ and $A$ was fixed by setting
the R-charge of a holomorphic four-form associated to the cone over $X
$ to two.
This four-form has a dual field theory interpretation as a
superpotential.
The relation between $A$ and $\psi$ fixes the normalization of the
gauge field action.

In addition to this perturbed metric, the RR four form $F_4$ is also
perturbed by $A$:
\be
F_4 = \frac{3r^2}{L^3} d^3 x \wedge dr - 4 L^3 ({\star_4}dA )\wedge
\omega \ .
\ee
Here $d^3x = dx^0 \wedge dx^1 \wedge dx^2$, and $\star_4$ is the
Hodge dual in
the AdS$_4$ directions only.
With $A=0$, $F_4$ can be thought of as the electric
flux from a stack of M2-branes spanning the $x^i$ coordinates.

With these formulae for $F_4$ and $ds^2$ in hand, we can normalize
the gauge field.
The 11 dimensional supergravity action is
\be
\frac{1}{2\kappa^2} \int d^{11} x \sqrt{-g} R -
\frac{1}{4\kappa^2} \int \left( F_4 \wedge \star F_4 + \frac{1}{3}
A_3 \wedge F_4 \wedge F_4 \right) \ .
\label{Seleven}
\ee
The first two
terms both give contributions to $|F|^2$, where $F=dA$.  In
particular, in making $A$ nonzero, the Ricci scalar becomes
\be
R = \tilde R - \frac{L^2}{4} |F|^2 + \frac{21}{2L^2}\ ,
\label{Rexp}
\ee
where $|F|^2 = F^{AB} F_{AB}$ and $\tilde R$ is the scalar curvature
in the $AdS_4$ directions.
Meanwhile, the four form produces a term of the form
\be
F_4 \wedge \star F_4 = -\left( \frac{9}{L^2}+\frac{3}{2} L^2 |F|^2
\right) \sqrt{-g}\; d^{11} x \ .
\label{Fnorm}
\ee

We cannot simply reduce the eleven dimensional action to an effective
four dimensional action
as can be seen from the form of $R$ and $|F_4|^2$.  Combining (\ref
{Rexp}) and (\ref{Fnorm})
in (\ref{Seleven}) leads to a Maxwell term $|F|^2$ of the wrong sign.
The reason Kaluza-Klein reduction does not commute with computing the equations of motion is related to the fact that the Bianchi identity $d F_4 = 0$ imposes the equation of motion $d\star F = 0$ on the gauge field.

Instead, we must reduce the eleven dimensional equations of motion
and from the
effective four dimensional equations of motion reconstruct a four
dimensional action.
Along with Maxwell's equations for $F$,
the eleven dimensional equations of motion reduce to
\be
R_{MN} = 2 L^2 \left( F_M{}^P F_{NP} - \frac{1}{4} g_{MN} |F|^2
\right) -
\frac{3}{L^2}g_{MN}
\ ,
\ee
which can be obtained from the four dimensional action
\be
S_{\rm{eff}}= \frac{1}{2\kappa_4^2} \int d^4x \sqrt{-g}  \left(\tilde
R - L^2  F_{MN} F^{MN}
+ \frac{6}{L^2} \right) \ .
\ee

Assuming that the four and eleven dimensional gravitational couplings
are
related by the volume of the compact manifold $X$,
\[
\frac{1}{2 \kappa_4^2} = \frac{(2L)^7 \Vol(X) }{2 \kappa^2} \ ,
\]
and using the standard normalization for $\kappa$ (\ref{kelevennorm}),
we find that the action for the gauge field becomes
\be
- \frac{\sqrt{2} N^{3/2}}{2^3\pi^5} \Vol(X) \int d^4x \sqrt{-g_4} |F|
^2 \ .
\ee
The volume of a seven sphere is $\Vol(S^7) = \pi^4 / 3$.

While in the case of more highly symmetric spaces, the R-symmetry
gauge field
transforms under a larger group,
based on the underlying Sasaki-Einstein structure, this U(1) subgroup
is in some
sense the most geometrically natural.

We conclude this section by explaining, for the case of $S^7$,
which $U(1)$ subgroup of $SO(8)$ we have extracted.  Earlier,
we stated that the $U(1)$ is normalized in reference to a holomorphic
four-form on the cone over the Sasaki-Einstein space.  For $S^7$,
the cone is $\mathbb{C}^4$, and the four-form
$\Omega = dX_1 \wedge dX_2 \wedge dX_3 \wedge dX_4$ where the $X_a$
are complex coordinates on $\mathbb{C}^4$.  Giving $\Omega$ R-charge
two means each $X_a$ will have R-charge one half and will transform
under
the $U(1)$ group action as $X_a \to e^{i \alpha /2} X_a$ for some phase
angle $\alpha$ which runs from zero to $2\pi$.

The Lie algebra for $SO(8)$ has four generators $\lambda_a$
in its Cartan sub-algebra which
we can choose to act on the $X_a$
as $\exp(i \alpha \lambda_a)(X_b) = \delta_{ab} e^{i \alpha/2} X_a$.
With this
normalization, $\tr \lambda_a \lambda_b = \frac{1}{2} \delta_{ab}$.
Comparing with the action of our special $U(1)$ subgroup, we see
that our $U(1)$ Lie algebra element $\lambda$ is a sum of the $
\lambda_a$:
$\lambda = \sum_a \lambda_a$.
Thus, $\tr \lambda^2 = 2$.

\section{Analytic solution}
\label{app:soln}
\noindent
At zero momentum, the mode equation (\ref{eq:Ay})
for $M(u)\equiv M_{yy}(u)$ takes the form
\begin{equation}
     f(u) \partial_u [f(u)\partial_u M(u)] + \wn^2 M(u)=0\,,
\end{equation}
with $f(u){=}1{-}u^3$.
By introducing a new coordinate
$z{=}\int_0^u d\tilde u/f(\tilde u)$, the equation simplifies,
\begin{equation}
    \partial_z^2 M(z) + \wn^2 M(z) = 0\,,
\end{equation}
with the boundary condition $M(z{=}0)=1$ at the boundary,
and the outgoing condition at the horizon $z{=}\infty$.
The solution is
\begin{equation}
    M(z) = e^{i\wn z}\,.
\end{equation}
That it corresponds to outgoing waves can be seen from the fact
that in the function $e^{-i\omega(t-z)}$ the wave front moves
toward larger $z$, i.e. closer to the horizon as $t$ increases.
Therefore we find
\begin{equation}
    M(u) = \exp\left[i\wn \int_0^u \frac{d\tilde u}{f(\tilde u)}
\right] \ .
\end{equation}
The leading asymptotics for $u$ near zero is
$M(u)=1+i\wn u$.
  From the AdS/CFT prescription (\ref{C-prescr}) we immediately find
\begin{equation}
   C_{yy}(\wn,0)=\Pi^T(\wn,0)=\Pi^L(\wn,0) = -i\chi\wn\,.
\end{equation}
This agrees with the result for conductivity
in section \ref{sec:conductivity}, as it should.

\section{Gauge field with a dilaton}
\label{app:d2}
\noindent
Consider a U(1) gauge field on a four dimensional manifold $M$
with an action of the form
  \be
  S = -\frac{1}{2 \gfourD^2} \int_M e^{-2 \phi} F \wedge \star F \ .
  \label{action}
  \ee
  There are a number of interesting 2+1 dimensional field theories
which have a dual
  R-symmetry gauge field of this type -- for example the M2-brane
theory at finite R-charge
  chemical potential and the D2-branes in type IIA string theory.
Here $F$ is the two-form gauge field, $\phi$ a dilaton like scalar,
and $\gfourD$ the coupling.
  The Maxwell equations can be written elegantly as
  $ d F = 0$ and $d \star e^{-2 \phi}  F = 0$.
  There is an equivalent S-dual theory where the roles of $F$ and $
\widetilde{F} \equiv \star e^{-2 \phi} F$
  are interchanged and we send
  $\gfourD e^\phi \to \widetilde{g}_{\mathrm{4D}} e^{\widetilde
{\phi}} \equiv 1/ (\gfourD e^\phi)$:
  \be
   S = -\frac{1}{2 \widetilde{g}_{\mathrm{4D}}^2} \int_M e^{2 \phi}
\widetilde{F} \wedge \star \widetilde{F} \ .
  \ee
The point we would like to emphasize is that when $\phi$ is a
constant, the theory is almost
self-dual in the sense that the equations of
motion for $F$ and $\widetilde{F}$ are identical.  When $\phi$ is not
a constant, the equations
of motion for $F$ and $\widetilde{F}$ are identical up to sending $
\phi \to -\phi$.


We would like to investigate the consequences of this duality in the
context of the AdS/CFT correspondence
where this gauge field is interpreted as a bulk field corresponding
to some global U(1) symmetry
on a $2{+}1$ dimensional boundary theory.
To this end, we assume the metric takes the diagonal form
  \be
  ds^2 = -g_{tt}(u) dt^2 + du^2 + g_{xx}(u) (dx^2 + dy^2) \ ,
  \ee
  where the metric components are only radially dependent on a
coordinate
  we call $u$.  By diffeomorphism invariance, we can always set $g_
{uu} = 1$.
  The boundary is taken to be located at $u=0$ and the interior for
$u>0$ with
  a horizon at $u=u_h > 0$.
  We will assume
  that as $u \to 0$, $-g_{tt} \sim g_{xx} \sim c^2/u^{\alpha}$ where $
\alpha > -2$.

We will calculate two-point functions of the U(1) current $J$
corresponding to this global symmetry.
Introducing a vector potential $F = dA$,
the retarded two-point function can be found using the method
described in
Sec.~\ref{sec:ads-cft}.  Namely, one looks for the solution to the
field equation for $A_\nu$ of the form
%
vector potential of the form
\be
A_\nu(x,t,u) = e^{i p \cdot x} {M_\nu}^\mu(p,u) A^0_\mu(p) \ ,
\ee
where ${M_\nu}^\mu(p,u)$
satisfies the radial component of the equation of motion for $A_\nu$.
Furthermore,
${M_\nu}^\mu(p,0)  = \delta_\nu^\mu$ and ${M_\nu}^\mu$ satisfies
outgoing boundary condition at the horizon $u=u_h$.
If the kinetic term for $A_\mu$ can be written as
%
\be
-\frac{1}{2\gfourD^2} \int du \, d^3x \, G(u) (A_\mu')^2
\label{kinetic}
\ee
then
\be
C^{\mu\nu}(k) =  -\frac{1}{\gfourD^2} \lim_{u\to 0} G(u) \frac
{\partial}{\partial u} M^{\mu \nu}(p,u)\ .
\ee

In particular, we take $A^\mu(x,t,u)$ to satisfy the equation of motion
\be
\partial_A \left[ e^{-2\phi} \sqrt{-g} g^{A B} g^{C D} (A_{B,D} - A_
{D,B})\right]=0 \ ,
\label{myeom}
\ee
and fix a radial gauge $A_u = 0$.  We also choose $p^\mu = (\omega, k,
0)$.
In this gauge, the equation of motion for $A_y$ becomes
\be
\partial_u \left[ e^{-2\phi} \sqrt{-g} g^{xx} A_y' \right] - k^2 \sqrt
{-g} (g^{xx})^2 e^{-2\phi} A_y - \omega^2 \sqrt{-g} g^{xx} g^{tt} e^
{-2\phi} A_y = 0 \ .
\label{Ayeom}
\ee
Because of the constraint on $\alpha$, the near boundary behavior of
$A_y$ ($u \sim 0$)
is governed by an expansion of the form
\be
M^{yy}(p,u)  = (1 + {\mathcal{O}}(u))  + u^{1+\alpha/2} {\mathcal{B}}
(1+ {\mathcal{O}}(u)) \ .
\ee
In the case where $\alpha$ is an even integer, the two series will
overlap, leading
to logarithmic terms in the first series, which complicate the story
but should not
alter it in any fundamental way.
The constant ${\mathcal B}$ is a complicated function of $\omega$ and
$k$ which
is determined by fixing outgoing boundary conditions at the horizon
$u=u_h$.
For $A_y$, the function $G(u)$ in (\ref{kinetic}) is
$\sqrt{-g} g^{xx} $ which near the boundary scales as $ c u^{-\alpha/
2}$ where
$c$ depends on the precise form of our metric.
By absorbing $\phi(0)$ into the value of $\gfourD$, we can
choose $\phi(0) = 0$.
  From this expansion and the form of $G(u)$, clearly
\be
C^{yy} = \frac{1}{ \gfourD^2} (1+ \alpha/2) c {\mathcal{B}} \ .
\ee

In our gauge, $A_y$ can be reinterpreted as a radial magnetic field,
$B_u = F_{xy} = - i k A_y$.
By electric-magnetic duality, replacing $\phi$ with $-\phi$,
the equation of motion for $B_u$ (\ref{Ayeom}) must be the same
as the equation of motion for $E_u \equiv -(\star e^{-2\phi} F)_{xy}
= \sqrt{-g} g^{tt} e^{-2\phi} A_t'$:
\be
\partial_u \left[ e^{2\phi} \sqrt{-g} g^{xx} E_u' \right] - k^2 \sqrt
{-g} (g^{xx})^2 e^{2\phi} E_u - \omega^2 \sqrt{-g} g^{xx} g^{tt} e^{2
\phi} E_u = 0 \ .
\label{Eueom}
\ee
We thus know
that $E_u$ has the near boundary expansion
\be
E_u = E_u^0 e^{i p \cdot x} \left[ (1 + {\mathcal{O}}(u))  + u^{1+
\alpha/2}
\widetilde{\mathcal{B}}(1+ {\mathcal{O}}(u)) \right] \ .
\ee
The tilde over ${\mathcal B}$ indicates it was derived from (\ref
{Ayeom}) having replaced
$\phi$ with $-\phi$.  In the case $\phi= \mbox{const}$, ${\mathcal B}
= \widetilde{\mathcal B}$.
We now use Gauss's law to constrain the boundary behavior of $E_u^0
$.  The equation
of motion following from taking the index $C = t$ in (\ref{myeom}) is
\be
\left( \sqrt{-g} g^{tt} e^{-2\phi} A_t' \right)' - k \sqrt{-g} g^{tt}
g^{xx} e^{-2\phi} ( \omega A_x + k A_t) = 0 \ .
\ee
  From the near boundary behavior, we find that
\be
E_u^0 (1+ \alpha/2)  \widetilde {\mathcal B} = -\frac{k}{c}  (\omega
A_x^0 + k A_t^0)  \ .
\ee

We can run a similar analysis of the component $A_x'$ and construct a
full boundary action.
We find that
\begin{eqnarray}
S_b &=& - \frac{1}{2 \gfourD^2} \int_{\mathbb{R}^3} d^3x \left[
{\mathcal B} c (1 + \alpha/2) (A_y^0)^2 -  \right.  \nonumber \\
&&
  \left. \frac{1}{\widetilde {\mathcal B} c (1+ \alpha/2)} \left(
k^2 (A_t^0)^2 + \omega^2 (A_x^0)^2 + \omega k A_t^0 A_x^0
\right)
\right]
\ .
\end{eqnarray}
  From this normalization, we conclude that the remaining two point
functions are
\begin{eqnarray}
C^{tt} &=& -\frac{1}{ \gfourD^2} \frac{k^2}{c (1+\alpha/2) \widetilde
{\mathcal B}}  \ , \\
C^{xt} &=& -\frac{1}{\gfourD^2} \frac{k\omega}{c(1+\alpha/2)
\widetilde {\mathcal B}} \ , \\
C^{xx} &=& -\frac{1}{\gfourD^2} \frac{\omega^2}{c(1+\alpha/2)
\widetilde {\mathcal B}} \ .
\end{eqnarray}
In the special case where $\phi$ is a constant and hence $\widetilde
{\mathcal B} = {\mathcal B}$,
we find that
\be
C^{tt} C^{yy} = -\frac{1}{ \gfourD^4} k^2 \; ; \; \;
C^{xt} C^{yy} = -\frac{1}{ \gfourD^4} k \omega \; ; \; \;
C^{xx} C^{yy} = -\frac{1}{ \gfourD^4} \omega^2 \ .
\ee

\end{document}